\documentclass[twocolumn]{aastex63}

\usepackage{apjfonts}

\def\d3{$\delta_{3}$ }
\def\1d3{$(1 + \delta_{3})$ }
\def\l1d3{$\log_{10}(1 + \delta_{3})$ }

\def\s3{$\Sigma_{3}$}

\def\24m{24 $\mu$m}

\def\kms{${\rm km~s^{-1}}$ }

\def\h2{$\rm H_{2}$}

\def\Mh2{$\rm M_{H_{2}}$}

\def\sigh2{$\Sigma_{\rm H_{2}}$}

\def\fh2{$f_{\rm H_{2}}$}

\shortauthors{Ho et al.}

\begin{document}

\title{Inside-Out vs. Outside-In Quenching of MaNGA Galaxies: Dependence on Stellar Mass and Environment}

\author{Zi-Hua Ho}
\altaffiliation{Email: zhho@asiaa.sinica.edu.tw}
\affiliation{Institute of Astronomy, National Tsing Hua University, Hsinchu 30013, Taiwan}
\affiliation{Institute of Astronomy \& Astrophysics, Academia Sinica, Taipei 10617, Taiwan}

\author{Lihwai Lin}
\affiliation{Institute of Astronomy \& Astrophysics, Academia Sinica, Taipei 10617, Taiwan}

\author{Hung-Yu Jian}
\affiliation{Institute of Astronomy \& Astrophysics, Academia Sinica, Taipei 10617, Taiwan}

\author{Bau-Ching Hsieh}
\affiliation{Institute of Astronomy \& Astrophysics, Academia Sinica, Taipei 10617, Taiwan}

\author{Carlos L\'{o}pez-Cob\'{a}}
\affiliation{Institute of Astronomy \& Astrophysics, Academia Sinica, Taipei 10617, Taiwan}

\author{S. F. Sánchez}
\altaffiliation{Email: sfsanchez@astro.unam.mx}
\affiliation{Instituto de Astronomía, Universidad Nacional Autónoma de México, A.P. 70-264, C.P. 04510, México, D.F., Mexico}

\author{Wen-Yen Wu}
\affiliation{Department of Physics, National Taiwan Normal University, No. 88, Section 4, Tingzhou Road, Taipei 116, Taiwan}
\affiliation{Institute of Astronomy \& Astrophysics, Academia Sinica, Taipei 10617, Taiwan}

\author{Shuai Feng}
\altaffiliation{Email: sfeng@hebtu.edu.cn}
\affiliation{College of Physics, Hebei Normal University, 20 South Erhuan Road, Shijiazhuang 050024, People's Republic of China}

\author{Shiyin Shen}
\altaffiliation{Email: ssy@shao.ac.cn}
\affiliation{Key Laboratory for Research in Galaxies and Cosmology, Shanghai Astronomical Observatory, Chinese Academy of Sciences, 80 Nandan Road, Shanghai 200030, People's Republic of China}

\begin{abstract}
    Galaxy quenching, the cessation of star formation, can proceed in spatially distinct ways, commonly described as inside-out or outside-in. However, the inferred quenching pattern depends strongly on how quenched or quenching regions are defined observationally. We utilize a sample of $\sim10{,}000$ galaxies from the Mapping Nearby Galaxies at APO (MaNGA) DR17 survey to systematically compare four widely used diagnostics of star formation suppression—specific star formation rate (sSFR), the 4000~$\mathring{A}$ break (D$_n$4000), post-starburst (PSB), and low-ionization (nuclear) emission-line region (LI(N)ER) emission—to examine how tracer choice influences the inferred spatial quenching pattern. Using the non-parametric method developed by \citet{lin19}, we classify galaxies into inside-out and outside-in quenching modes based on the location on the plane of the fraction of the quenched area (Fq) and the concentration of quenched area (Cq). We find that sSFR criterion yields comparable proportions of galaxies classified as inside-out and outside-in, while D$_n$4000 and LI(N)ER diagnostics strongly favor inside-out patterns. Because PSB traces a distinct transitional phase, PSB-selected spaxels occupy a different region of the Fq-Cq plane. Across most diagnostics, the fraction of galaxies classified as inside-out increases with stellar mass, while outside-in patterns are more common in lower-mass systems, especially among satellites. In contrast, the dependence of quenching mode on halo mass is weaker and less consistent across diagnostics. These differences show that the tracers probe complementary stages and timescales of star-formation suppression, and together provide a more complete view of spatially resolved quenching.
\end{abstract}


\section{introduction}
\label{sec:introduction}
Galaxies are traditionally classified into two distinct populations: the “blue cloud” and the “red sequence,” as observed in the color-magnitude diagram \citep[e.g.,][]{str01,bal04,bel04,fab07,bla09,ilb10,wet12,muz13}. Similarly, in the context of the stellar mass-star formation rate (SFR) diagram, galaxies are categorized as either star-forming, which follow the star-forming main sequence (SFMS; e.g., \citep{bri04,dad07,elb07,noe07,pan09,elb11,lin12,lin14,whi12,ren15,boo18}), or as passive, “quenched” galaxies. An additional, less common population, known as “green valley” galaxies, occupies the intermediate space between the SFMS and quenched galaxies and is considered to represent a transitional phase. As star-forming galaxies cease their star formation, they migrate away from the SFMS, traverse the green valley, and ultimately settle into the red sequence \citep[e.g.,][]{fab07,mar07,sal07,sch07,wyd07,men11,sch14,sme15,mah17,bel18}.

A variety of processes have been proposed to be responsible for quenching star formation in galaxies \citep[e.g., see the review by][]{man18}. For example, galaxy mergers can induce a brief starburst, consuming available gas and leading to a rapid cessation of star formation \citep[mergers;][]{sch14,ell22,ell24}. Additionally, the transition of dark matter halos from a fast accretion phase to a slow accretion phase reduces the availability of cold gas \citep[cosmological starvation;][]{fel15}. The collapse of gas inflow under certain conditions can result in heating, preventing the supply of cold gas or inhibiting gas cooling \citep[virial shock;][]{bir03}. Stellar feedback, such as supernovae (SNe), can also expel gas from the galaxy \citep[SNe feedback;][]{mur15,hop18,li20}. Furthermore, the activity of active galactic nuclei (AGN) is theorized to regulate star formation by heating the gas and expelling cold gas, as supported by various theories and simulations \citep[AGN feedback;][]{di05,spr05,fab12,blu16,hop16}. 

However, each theoretical quenching mechanism is associated with different timescales, stellar masses, halo masses, and other physical properties of galaxies \citep[][]{sme15,jia20,ham23}. Additionally, galaxy quenching can result from multiple simultaneous mechanisms, and one mechanism may trigger another \citep[][]{san21}.

While observationally it is challenging to disentangle different quenching processes directly, some insights can be gained from having spatially resolved observations. Depending on the sequence of quenching, galaxy quenching modes can be broadly classified as inside-out \citep[][]{per13,li15,lop18,lin19,blu20,pap22} or outside-in \cite[][]{koo04}. Quenching mechanisms driven by internal or central processes, including morphological quenching and AGN feedback, tend to suppress star formation in the inner regions and have less impact in the outskirts \citep{guo19,blu20}. In contrast, environmentally driven mechanisms, including ram pressure stripping and tidal interactions, preferentially remove or heat gas in the outer regions, leading to an outside-in quenching pattern \citep[][]{gun72,koo04}.

As integral field spectroscopy (IFS) has advanced, the technical definitions of quenching have become increasingly diverse across studies. Some studies rely on a single parameter to define quenched regions. For example, \citet{hong23} uses the 4000 Angstrom break (D$_n$4000 index), an age indicator of the stellar population, to classify spaxels as quenched or star-forming. Others, such as \citet{col20}, \citet{ell21a}, \citet{ell21b}, \citet{kal22}, and \citet{ar23}, use the H$\alpha$ equivalent width (EW(H$\alpha$)) while \citet{rat22} employ the specific star formation rate (sSFR). Some studies use a combination of parameters: \citet{cc23} use a dividing line based on both the D$_n$4000 index and EW(H$\alpha$) to classify aging and quenched regions, \citet{cc21} use EW(H$\alpha$) and the color index (g - r) to define regions at different quenching stages, and \citet{blu20} use the deviation of SFR surface density ($\Delta \Sigma_{\mathrm{SFR}}$) from the SFMS to identify quenched regions. Additionally, some studies employ more complex and specific definitions: \citet{chen19} and \citet{cheng24} use the H$\delta$ absorption line (H$\delta_{\rm A}$) and EW(H$\alpha$) to select post-starburst (PSB) galaxies as quenching galaxies and then use the radial profile of D$_n$4000, H$\delta_{\rm A}$ and EW(H$\alpha$) to determine the quenching mode, while \citet{lin19} combines the low-ionization (nuclear) emission-line region (LI(N)ER) in the Baldwin–Phillips–Terlevich (BPT) diagnosis diagram \citep[][]{bal81} with EW(H$\alpha$) to define the quenched region. Each of these approaches provides unique insights into the processes driving galaxy evolution.

The lack of a unified definition of galaxy quenching, together with the use of diverse selection methodologies, leads to substantial differences in how quenched regions are identified in integral field unit (IFU) studies. As a result, the inferred quenching state or spatial quenching mode of a given galaxy can depend strongly on the specific tracer or criterion adopted. Relying on a single selection method applied to a specific galaxy population may therefore provide an incomplete or biased view of the quenching process. In this study, we utilize a sample of 10220 galaxies from the Mapping Nearby Galaxies at APO (MaNGA) survey \citep[][]{bun15} to compare spatial quenching modes using four different tracers: sSFR, D$_n$4000, PSB, and LI(N)ER. By adopting two non-parametric parameters introduced by \cite{lin19}—$\mathrm{F}_q$ (the fraction of quenched area) and $\mathrm{C}_q$ (the concentration of quenched area)—we are able to investigate quenching modes in greater detail, particularly in galaxies exhibiting spatially patchy quenching. 

In \S\ref{sec:data}, we describe the data and catalogs used in our analysis, including the quantities from Pipe3D product from MaNGA DR17 and the halo mass catalog from Yang's group catalog \citep{Yang05,Yang07}. In \S\ref{sec:method}, we introduce the method for analyzing galaxies’ quenching modes, specifically distinguishing between inside-out and outside-in quenching, following the approach of \citet{lin19}. In \S\ref{sec:results}, we present the main results concerning the inside-out vs. outside-in quenching modes based on the four different tracers of quenched areas. Discussions and Conclusions are given in 
\S\ref{sec:discussion} and \S\ref{sec:conclusion}, respectively.

Throughout this paper we adopt the following cosmology: \textit{H}$_0$ = 70~\kms Mpc$^{-1}$, $\Omega_{\rm m} = 0.3$ and $\Omega_{\Lambda } = 0.7$. We use a Salpeter initial mass function (IMF).

\section{data}
\label{sec:data}
%
\subsection{MaNGA DR17}
\label{subsec:MaNGA DR17}
The MaNGA survey employs integral field spectroscopy to observe galaxies in the nearby universe \citep[][]{bun15,yan16a,yan16b,wak17}. In this study, we analyze 10,220 galaxies with $z < 0.15$ from the MaNGA data in Sloan Digital Sky Survey (SDSS) data release 17 (DR17) \citep[][]{san22}. The IFS observations provide three-dimensional data cubes, with two spatial dimensions and one spectral dimension, enabling spatially resolved studies of galaxy properties.

We primarily use the publicly available Pipe3D value-added catalog (VAC) for MaNGA DR17 \citep[][]{san22}, which deliver spatially resolved stellar population and emission-line properties.
The Pipe3D pipeline performs stellar population synthesis fitting and emission-line measurements on MaNGA data using simple stellar population (SSP) models from the \texttt{MaStar\_sLOG} library, which is based on the MaNGA stellar library (MaStar) and covers a broad range of stellar atmospheric parameters \citep[][]{yan19,san22}. All spaxel-based spectroscopic quantities used in this work are taken directly from the Pipe3D DR17 VAC.

For this study, we focus on a subset of spatially resolved quantities from the Pipe3D DR17 VAC relevant to identifying and characterizing quenched regions. These include two-dimensional maps of stellar mass surface density ($\Sigma_\ast$), the D$_n$4000 index, and emission-line fluxes of H$\alpha$, H$\beta$, [N\,\textsc{ii}]$\lambda6584$, and [O\,\textsc{iii}]$\lambda5007$, as well as the equivalent width of H$\alpha$. The SFR are derived from dust-corrected H$\alpha$ luminosities based on the extinction law with $R_V = 4.5$ \citep{cal01,fis05} and the calibration described in \citet{vog13}. The sSFR is computed by normalizing the SFR by the stellar mass.

\subsection{Criteria of quenched spaxels}
\label{subsec:Criteria of quenched spaxels}

\begin{figure}[th!]
    \centering
    \includegraphics[scale=0.155]{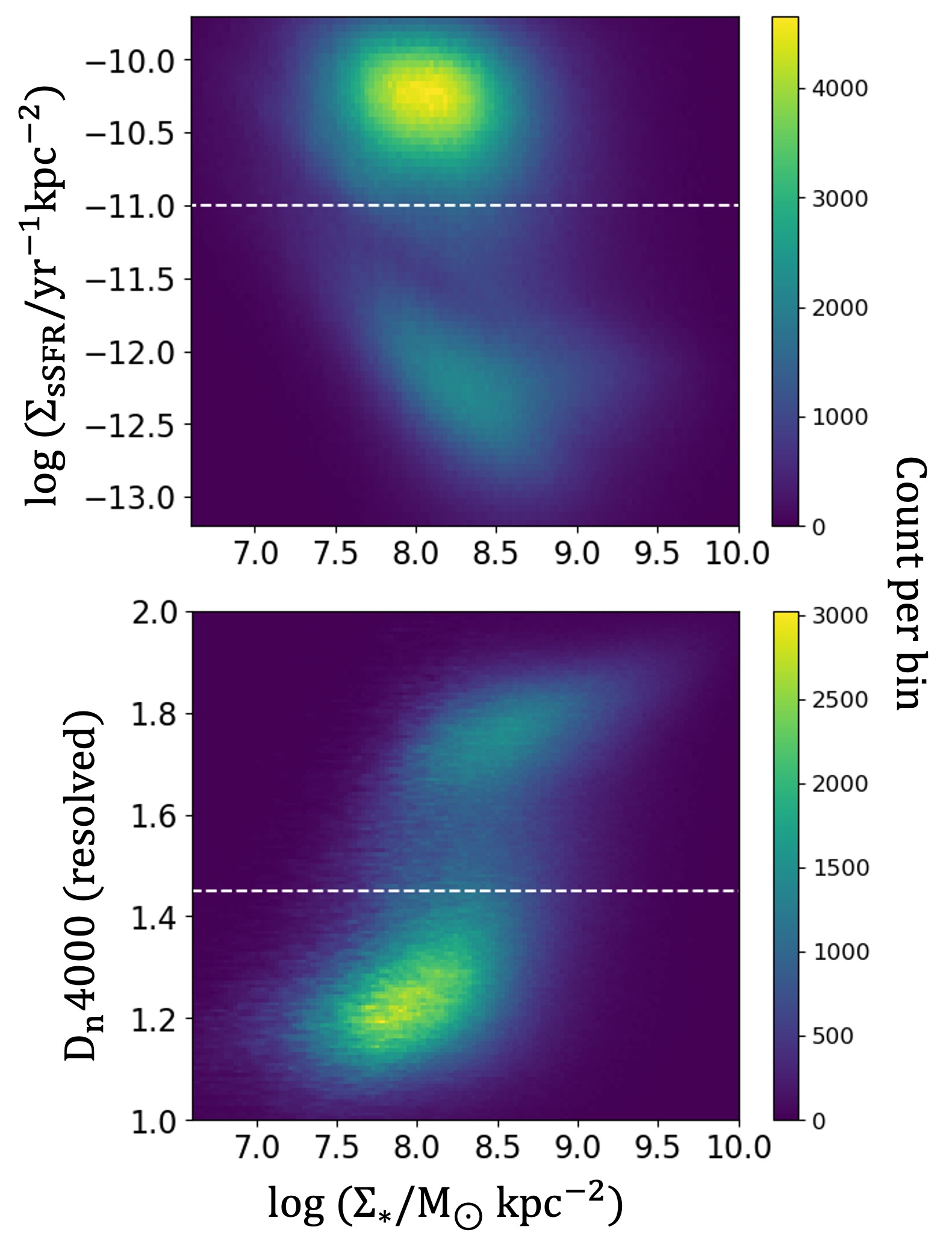}
    \caption{Distributions of sSFR (upper panel) and D$_n$4000 index (lower panel) for spaxels within 1.5 $R_{\text{e}}$. In both panels, the x-axis represents the stellar mass surface density ($\Sigma_*$) of the spaxels. The color scale indicates the number of spaxels per bin. The white dashed lines indicate the threshold values adopted in our criteria to separate quenched and non-quenched spaxels.}
    \label{fig:ssfrandd4000}
\end{figure}

\begin{figure*}[th!]
    \centering
    \includegraphics[scale=0.166]{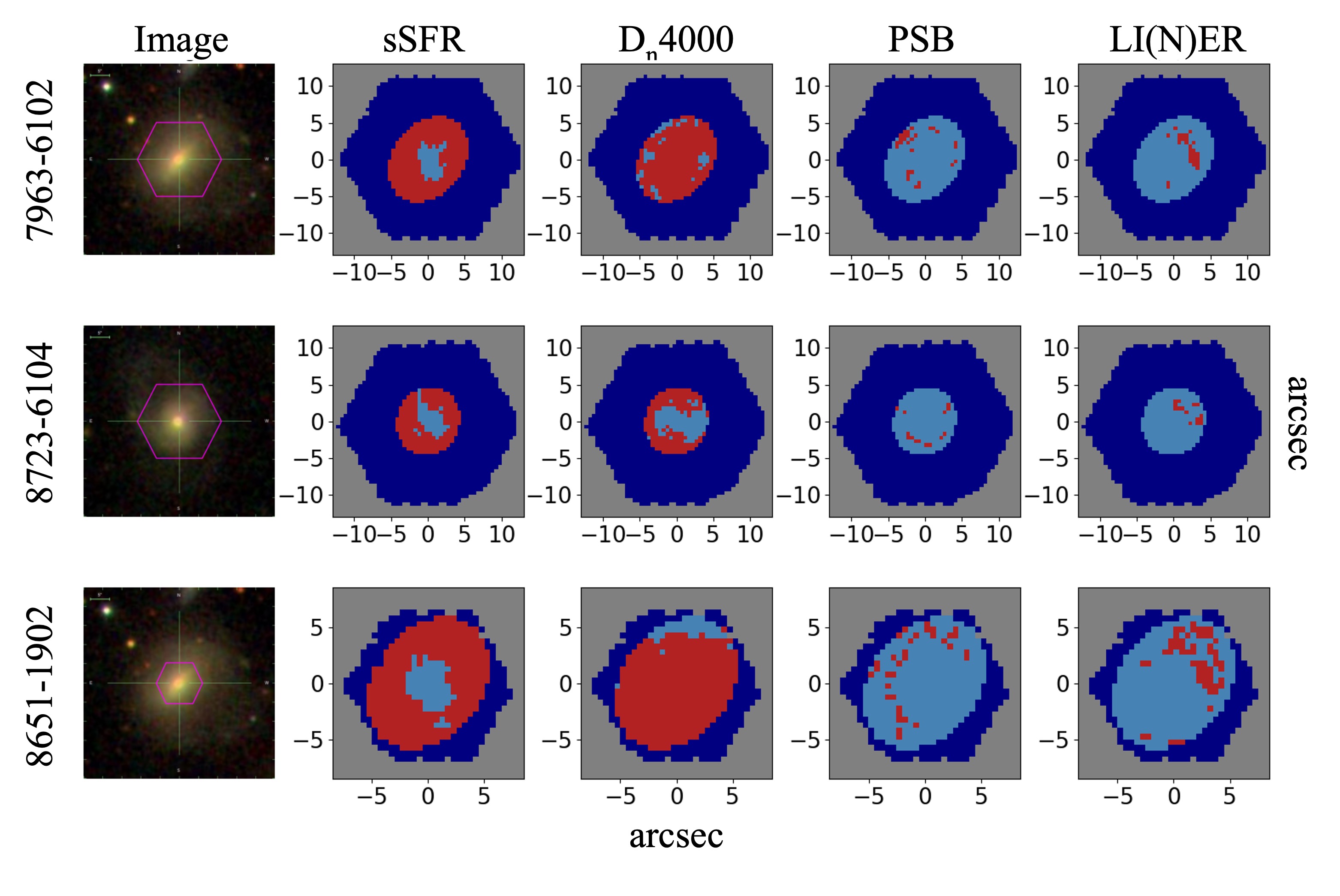}
    \caption{Maps of three representative galaxies illustrating the spatial distribution of spaxels selected by different quenching criteria. The leftmost column shows the SDSS 3-colour composite images of the galaxies, with the MaNGA field of view indicated by the hexagon. The remaining columns display the corresponding spaxel maps classified using the sSFR, D$_n$4000, PSB, and LI(N)ER criteria, respectively. In each spaxel map, red spaxels indicate regions classified as quenched by the corresponding criterion, except in the PSB panel, where they indicate quenching regions. Light blue spaxels represent spaxels that are not selected by that criterion (but are not necessarily star-forming). Dark blue spaxels denote regions that do not satisfy the common selection cuts on stellar mass surface density and galactocentric radius ($R_\mathrm{e}$), and grey spaxels indicate regions not covered by the MaNGA FoV.}
    \label{fig:maps}
\end{figure*}

\begin{figure*}[th!]
    \centering
    \includegraphics[scale=0.147]{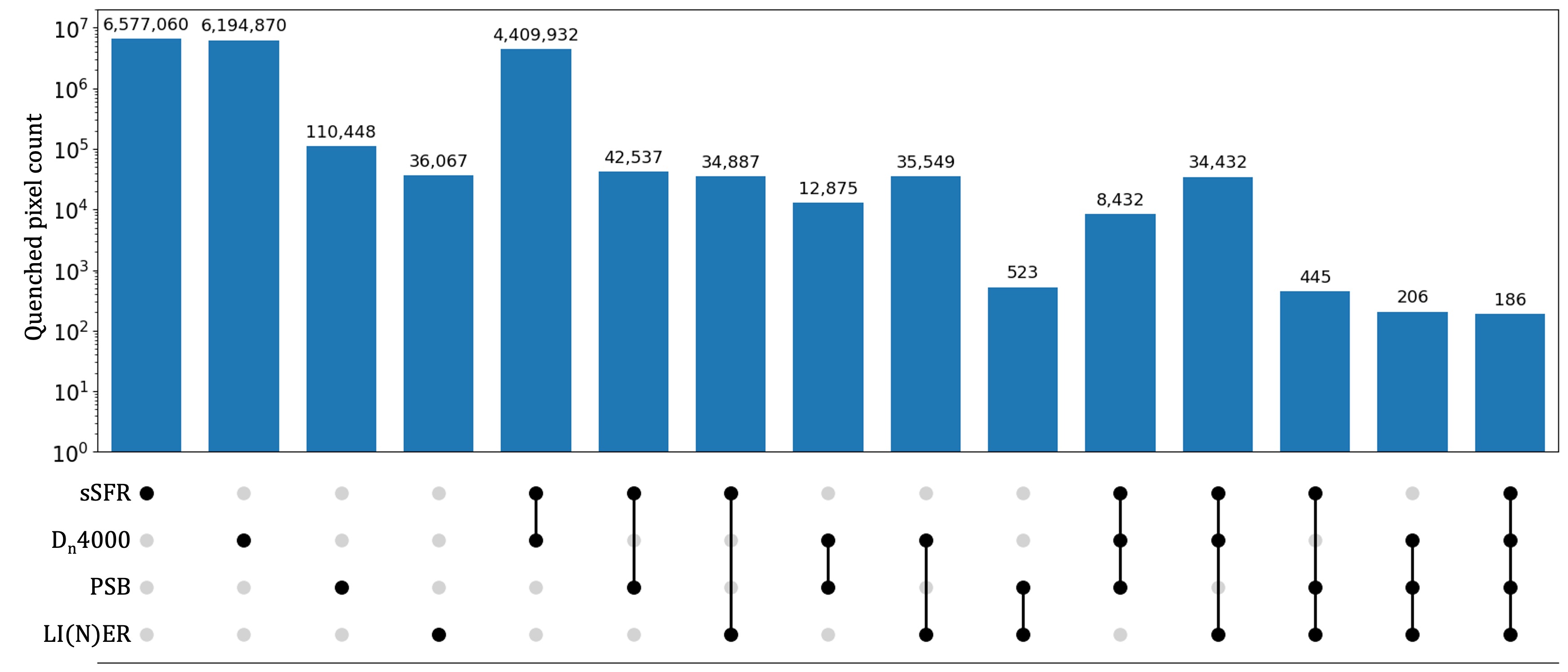}
    \caption{Spaxel-level overlap between four quenching diagnostics: sSFR, D$_n$4000, PSB, and LI(N)ER. Bars represent the total number of spaxels selected by each individual criterion as well as their inclusive intersections, where each combination includes all spaxels satisfying the corresponding set of criteria, regardless of whether additional criteria are also met. The matrix below indicates the corresponding combinations. The y-axis is shown in logarithmic scale to emphasize the wide dynamic range in spaxel counts.}
    \label{fig:upsetplot}
\end{figure*}

In this section, we detail the four quenching criteria used in our analysis. 

Before applying these criteria, we first impose a set of common spaxel-level selection cuts to ensure reliable measurements and uniform spatial coverage.

All spaxels included in the analysis are required to satisfy two conditions: (1) $\log (\Sigma_\ast/\mathrm{M}_\odot\,\mathrm{kpc}^{-2}) > 6.5$, and (2) to lie within 1.5 effective radii ($R_\mathrm{e}$) of the galaxy center. The stellar mass surface density threshold is adopted to exclude low-mass, low-surface-brightness spaxels, where stellar population and emission-line measurements are more uncertain. The radial limit of 1.5 $R_\mathrm{e}$ is adopted to ensure uniform radial coverage across the MaNGA sample. In the MaNGA survey, approximately two-thirds of galaxies are observed out to $\sim$1.5 $R_\mathrm{e}$, while the remaining one-third extend to $\sim$2.5 $R_\mathrm{e}$ \citep{bun15}. Adopting a common radial limit of 1.5 $R_\mathrm{e}$ therefore allows us to maximize the sample size while maintaining consistent spatial coverage.

To ensure the robustness of the galaxy-level classification, we further require that a galaxy contains at least six quenched spaxels in order to be identified as exhibiting quenching features. Increasing the minimum number of quenched spaxels reduces the total number of selected galaxies, but does not significantly change the dominant quenching mode or affect the overall conclusions of this study.

\medskip
\noindent\textbf{sSFR criterion.}
For the sSFR criterion, we adopt a threshold value of $\log (\Sigma_\mathrm{sSFR}/\mathrm{yr}^{-1}\,\mathrm{kpc}^{-2}) = -11$, such that spaxels with $\log (\Sigma_\mathrm{sSFR}/\mathrm{yr}^{-1}\,\mathrm{kpc}^{-2}) < -11$ are classified as quenched, while those above this threshold are considered non-quenched. This threshold is motivated by the sSFR distribution of our sample (Figure~\ref{fig:ssfrandd4000}), where adopting a higher value (e.g., $-10.7$) would result in significant overlap with the star-forming population and hinder a clean separation between quenched and star-forming spaxels. We therefore adopt a more conservative threshold of $-11$ to achieve a clearer division. We verify that this choice does not qualitatively affect our results: the fractions of inside-out and outside-in quenching both decrease by $\sim$3\%, while their relative dominance remains unchanged.

\medskip
\noindent\textbf{D$_n$4000 criterion.}
We adopt a threshold of D$_n$4000 $= 1.45$, such that spaxels with D$_n$4000 $> 1.45$ are classified as quenched. This choice is guided by the observed correlation between sSFR and D$_n$4000 in MaNGA data. Previous studies have shown that these two quantities are tightly correlated in star-forming regions, but the relation breaks down at D$_n$4000 $\gtrsim 1.45$ \citep{blu20,pan24}. In particular, \citet{pan24} demonstrated that D$_n$4000 $= 1.45$ corresponds approximately to $\log (\Sigma_\mathrm{sSFR}/\mathrm{yr}^{-1}\,\mathrm{kpc}^{-2}) \sim -10.7$ along the star-forming sequence. 

We further examine the D$_n$4000 distribution in our sample (Figure~\ref{fig:ssfrandd4000}) and find that, although the separation between populations is less pronounced than in sSFR, this threshold still provides a reasonable division between younger and older stellar populations. 

\medskip
\noindent\textbf{PSB criterion.}
For the PSB criterion, we adopt the selection defined by \citet{chen19}, which focuses on spectral features associated with A-type and F-type stars. A spaxel is identified as a PSB region if it satisfies H$\delta_{\text{A}} > 3 \mathring{A}$, EW(H$\alpha$) $< 10 \mathring{A}$, and $\log (\mathrm{EW(H}\alpha)/\mathring{A}) < 0.23 \times \mathrm{H}\delta_{\text{A}} - 0.46$, and we treat them as regions undergoing quenching in our analysis.

As illustrated by the star formation history (SFH) tracks presented in Appendix~\ref{app:sfh_timescales}, the PSB criterion traces a short-lived evolutionary phase associated with rapid or very recent quenching. Only SFHs undergoing rapid quenching enter the PSB selection window, and they remain within this region for a relatively brief time interval (Figure~\ref{fig:psb_track}). Consequently, spaxels that do not satisfy the PSB criterion may still be star-forming or may have been quenched for a longer time.

\medskip

\noindent \textbf{LI(N)ER criterion.}
For the LI(N)ER selection, we follow the approach of \citet{lin19}, classifying emission-line regions into star-forming, composite, LI(N)ER, and Seyfert regimes using the [N,\textsc{ii}]-BPT diagram \citep[][]{bal81,cf10}. We utilize four emission lines: H$\alpha$, H$\beta$, [N,\textsc{ii}], and [O,\textsc{iii}], with classifications based on the demarcation lines from \citet{kew01}, \citet{kau03}, and \citet{sta06}. Spaxels are required to have S/N $> 3$ for H$\alpha$ and H$\beta$, and S/N $> 2$ for [N,\textsc{ii}] and [O,\textsc{iii}]. LI(N)ER spaxels are identified within the [N,\textsc{ii}]-BPT diagram and are further classified as quenched regions if EW(H$\alpha$) $< 3 \mathring{A}$, following \citet{cid11} to distinguish ionization powered by evolved stellar populations from that associated with star formation or AGN activity.

The LI(N)ER phase is generally associated with evolved stellar populations and is understood to arise after the cessation of star formation, when ionization by hot low-mass evolved stars (e.g., post-asymptotic giant branch stars or HOLMES) becomes dominant. Stellar population models and observations suggest that such emission emerges after star formation has been largely suppressed, on timescales ranging from $\sim10^8$ yr to Gyr scales depending on the assumed star formation history and stellar evolution model, and can persist over extended periods as long as the underlying old stellar population remains in place \citep{bel16,byl19}.

In addition, the combination of BPT-based signal-to-noise requirements and the EW(H$\alpha$) threshold can introduce a selection bias. Since LI(N)ER emission from evolved stellar populations is intrinsically weak (typically EW(H$\alpha$) $\lesssim 3$ \AA), spaxels with the weakest emission may fail the S/N criteria required for BPT classification \citep{bel16}. As a result, the observed LI(N)ER population may be biased toward regions with relatively stronger emission, while genuinely weak LI(N)ER regions remain undetected. Our adopted definition therefore isolates a subset of BPT-classified LI(N)ER regions that is more closely associated with evolved, quenched stellar populations, rather than all regions classified as LI(N)ER based solely on emission-line ratios.

To provide a visual illustration of how these criteria select regions in practice, we present spatially resolved classification maps for three representative galaxies in Figure~\ref{fig:maps}. Each panel shows the distribution of spaxels selected by the sSFR, D$_n$4000, PSB, and LI(N)ER criteria, respectively. These maps highlight that, while the sSFR and D$_n$4000 criteria generally identify broadly similar quenched structures, the PSB and LI(N)ER selections trace more localized and distinct regions. In particular, PSB spaxels tend to appear in sparse and patchy configurations, consistent with their interpretation as regions undergoing recent or ongoing quenching, whereas LI(N)ER spaxels are more spatially extended and coherently distributed, tracing the underlying evolved stellar populations.

To further illustrate the relationship between these quenching criteria, we examine their spaxel-level overlap, as shown in Figure~\ref{fig:upsetplot}. We find that the sSFR and D$_n$4000 criteria identify broadly consistent quenched populations, with a substantial overlap of $\sim 4.4 \times 10^6$ spaxels. This reflects the well-known correlation between sSFR and stellar population age, although the two indicators probe slightly different aspects of the star formation history (see Appendix~\ref{app:sfh_timescales} for an illustrative comparison based on toy SFH models). As illustrated by those models, the relative ordering of the thresholds depends on the adopted diagnostic definitions: in most of the illustrative SFHs, the $D_n$4000 threshold is reached earlier than the sSFR threshold, whereas the sSFR criterion responds more strongly to variations in the quenching timescale. Thus, the two diagnostics overlap strongly, but do not trace identical stages of quenching.

The LI(N)ER-selected spaxels also overlap with both sSFR and D$_n$4000, but are expected to trace a more evolved phase. Literature studies suggest that LI(N)ER-like emission associated with evolved stellar populations typically emerges after star formation has been largely suppressed, on timescales of order $\sim 10^8$ yr to Gyr, and can persist for extended periods as long as the old stellar population remains in place \citep{cid11, yan12, bel16, byl19}.

By contrast, PSB-selected spaxels show only partial overlap with the sSFR- and D$_n$4000-selected populations, with the overlap being more evident for sSFR than for D$_n$4000. This is consistent with the idea that PSB traces a short-lived subset of rapidly quenching regions, rather than the full quenched or quenching populations identified by the broader sSFR and D$_n$4000 criteria.

This behavior is qualitatively consistent with an evolutionary sequence in which different tracers probe distinct timescales of quenching. PSB features correspond to a short-lived phase shortly after rapid quenching, while LI(N)ER-like emission, predominantly powered by HOLMES, emerges at later times after star formation has already ceased \citep[e.g.,][]{cid11,yan12,bel16,byl19}. In contrast, sSFR and D$_n$4000 provide broader coverage of quenched populations across a wider range of evolutionary stages. Overall, these results highlight that different quenching criteria probe complementary phases of galaxy evolution, and their combination provides a more comprehensive view of the quenching process.

The LI(N)ER criterion is not included in these SFH models, as it relies on emission-line diagnostics. Instead, its temporal sensitivity can be inferred from previous studies. A substantial fraction of LI(N)ER-like emission in galaxies has been shown to arise from ionization by hot low-mass evolved stars (HOLMES), particularly post-asymptotic giant branch (post-AGB) stars, rather than from ongoing star formation \citep[e.g.,][]{stasin08,cid11,bel16,byl19}. Such emission is typically associated with old stellar populations and can persist over extended periods, as it is tied to the underlying evolved stellar component. As a result, the LI(N)ER criterion preferentially selects regions in which star formation has already been largely suppressed, and therefore traces a relatively evolved or low-star-formation phase. Unlike the sSFR and $D_n$4000 criteria, which depend on the adopted thresholds and probe different aspects of star formation, the LI(N)ER selection is more directly associated with non-star-forming ionization sources and probes a later stage than the short-lived PSB phase.

\color{black}
\subsection{Halo mass catalog}
\label{subsec:Halo mass catalog}
We adopt the group catalog constructed by \cite{Yang07}, which has been widely used to characterize the galaxy environment in the local universe. This catalog is based on a halo-based group finder algorithm \citep{Yang05} applied to the galaxy sample of the SDSS DR7 \citep{aba09}. The algorithm begins with a friends-of-friends (FoF) search in redshift space to identify tentative group centers, then iteratively assigns galaxies to groups by estimating the underlying dark matter halo properties and maximizing the group membership likelihood. This procedure yields a physically motivated group membership assignment and enables the derivation of group halo masses through abundance matching. The catalog provides a range of parameters, including group halo mass, total luminosity, number of members, and the brightest group galaxy (BGG), offering a comprehensive view of the group-scale environment.

In this work, we use the PetroB version of the Yang catalog, which is constructed using Petrosian magnitudes and includes 472,113 galaxy groups with a total of 602,570 galaxies, all with SDSS spectroscopic redshifts. We cross-match this group catalog with the 10,782 galaxies in the MaNGA DAPALL catalog using a 2 arcsecond matching radius, resulting in 9,236 galaxies with group environment parameters. 

\section{method}
\label{sec:method}
To quantify the spatial distribution of quenched spaxels in a galaxy, we utilize the method defined by \citet{lin19}, which uses two non-parametric parameters to characterize quenching patterns. The first parameter, $\mathrm{F}_q$, describes the fraction of the quenched area within a galaxy:

\begin{equation}
    \mathrm{F}_q=\frac{N_{\text{quenched}}}{N_{\text{all}}}
\end{equation}

By definition, $\mathrm{F}_q$ ranges from 0 to 1. For convenience, we express $\mathrm{F}_q$ in units of percentage (i.e., $100 \times \mathrm{F}_q$) in figures and discussions, following the convention adopted in \citet{lin19}.

The second parameter, $\mathrm{C}_q$, represents the concentration of the quenched area:

\begin{equation}
    \mathrm{C}_q=\frac{\sum r_{\text{all}}^2}{\sum r_{\text{quenched}}^2}
\end{equation}

In these formulas, $N_{\text{quenched}}$ refers to the number of quenched spaxels, as defined in the previous section, and $N_{\text{all}}$ is the total number of spaxels that satisfy the common spaxel-level selection cuts, including the stellar mass surface density threshold, and lie within 1.5 $R_{\text{e}}$. The terms $r_{\text{quenched}}$ and $r_{\text{all}}$ represent the inclination-corrected distances from the galaxy center to the quenched spaxels and to all spaxels included in $N_{\text{all}}$, respectively.

In general, a higher $\mathrm{F}_q$ indicates a larger quenched area within the galaxy, while a higher $\mathrm{C}_q$ value suggests a more centrally concentrated quenched region.

In addition to these two parameters, $F_{qi}$ is introduced to distinguish between inside-out and outside-in quenching modes \citep[see Appendix in][for detailed derivation]{lin19}. $F_{qi}$ quantifies the contribution of the inner quenched area. Galaxies are classified as inside-out or outside-in quenched by adopting a threshold of $F_{qi}=50\%$ \citep[see the illustration in][Figure 2]{lin19}.

Galaxies with $F_{qi}>50\%$, meaning the inner quenched area dominates over the outer quenched area, are classified as inside-out quenched galaxies. Conversely, galaxies with $F_{qi}<50\%$ are classified as outside-in quenched. By utilizing these three parameters, we construct the $\mathrm{F}_q$-$\mathrm{C}_q$ plane to visualize the distribution of quenching modes and assess the relative contributions of inside-out and outside-in quenching.

Notably, this method is not well-suited for identifying the quenching pattern for fully quenched galaxies. Furthermore, it relies on the present-day spatial configuration of star formation, meaning that if the quenched regions were transient and have since evolved, they may no longer be detectable by this method. As a result, this approach may lead to an underestimation of the true number of galaxies that experienced spatially distinct quenching phases in the past.

\section{results}
\label{sec:results}
Motivated by the distinct physical timescales traced by different quenching indicators, particularly the PSB criterion, we construct two overlapping galaxy samples for comparison. Subsample A consists of galaxies that satisfy the sSFR, D$_n$4000, and LI(N)ER criteria, while Subsample B is a more restrictive subset that additionally satisfies the PSB criterion (i.e., sSFR + D$_n$4000 + PSB + LI(N)ER). The number of galaxies in each sample is summarized in Table~\ref{tab:sample_size}.

Before presenting the results, we clarify how we define galaxies selected by multiple quenching criteria. In this work, the term ’overlapping’ is defined at the galaxy level rather than on a spaxel-by-spaxel basis. A galaxy is considered to be selected by a given quenching criterion if it contains at least six spaxels classified as quenched by that criterion, following the definition in Section~\ref{sec:data}.
\color{black}
\begin{table}[h!]
\centering
\caption{Galaxy sample statistics.}
\label{tab:sample_size}
\begin{tabular}{cccc}
\hline\hline
 & Full Sample & Subsample A & Subsample B \\
\hline
Sample size & 10,220 & \quad 1,234 & \quad 172 \\
\hline
\end{tabular}
\end{table}

\subsection{Quenching Mode Fractions in the Full Sample}
\label{subsec:Quenching Mode Fractions in the Full Sample}
\begin{figure*}[tbh]
    \centering
    \includegraphics[scale=0.205]{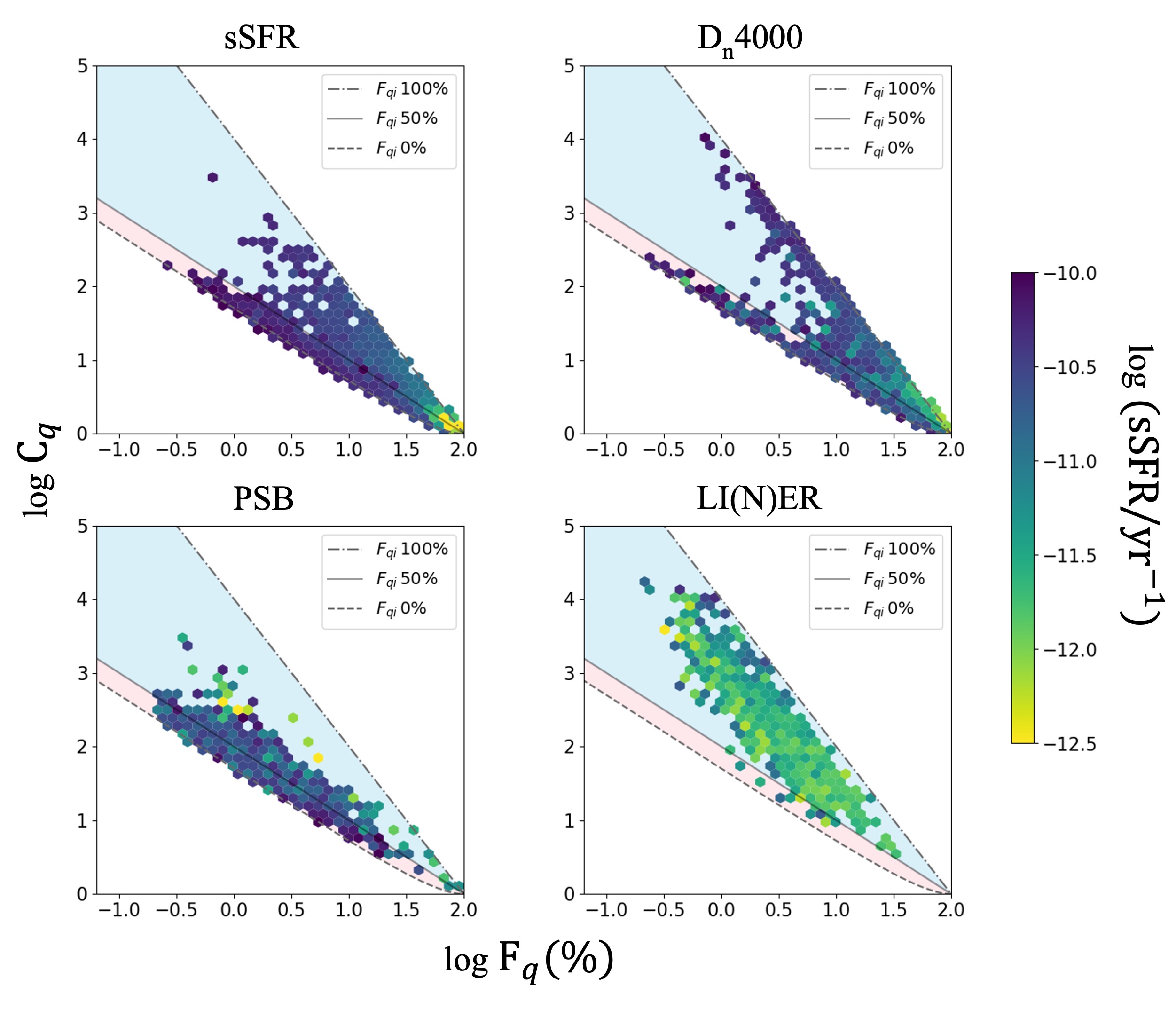}
    \caption{$\mathrm{F}_q$–$\mathrm{C}_q$ plane color-coded by the global $\log(\mathrm{sSFR}/\mathrm{yr}^{-1})$ of galaxies in the Full Sample (from left to right, top to bottom: sSFR, D$_{n}4000$, PSB, and LI(N)ER). The global sSFR values are taken from the Pipe3D DR17 catalog. The x-axis is $\mathrm{F}_q$, the fraction of selected area in units of percentage, such that higher $\mathrm{F}_q$ corresponds to a larger selected area, while the y-axis is $\mathrm{C}_q$, the concentration of the selected regions; lower $\mathrm{C}_q$ indicates a more centrally concentrated selected region. Given the large number of galaxies in the Full Sample, we adopt a hexagonal binning representation, in which each bin is color-coded by the mean global sSFR of galaxies falling within that region of the $F_q$–$C_q$ plane. Because the $F_q$–$C_q$ plane is geometrically bounded, some galaxies naturally accumulate near the edges of the allowed region. This figure is intended as a qualitative visualization of galaxy locations in the $F_q$–$C_q$ plane; quantitative quenching-mode fractions are shown in Figure~\ref{fig:barchart_full}.}
    \label{fig:fullsample}
\end{figure*}

\begin{figure}
    \centering
    \includegraphics[scale=0.125]{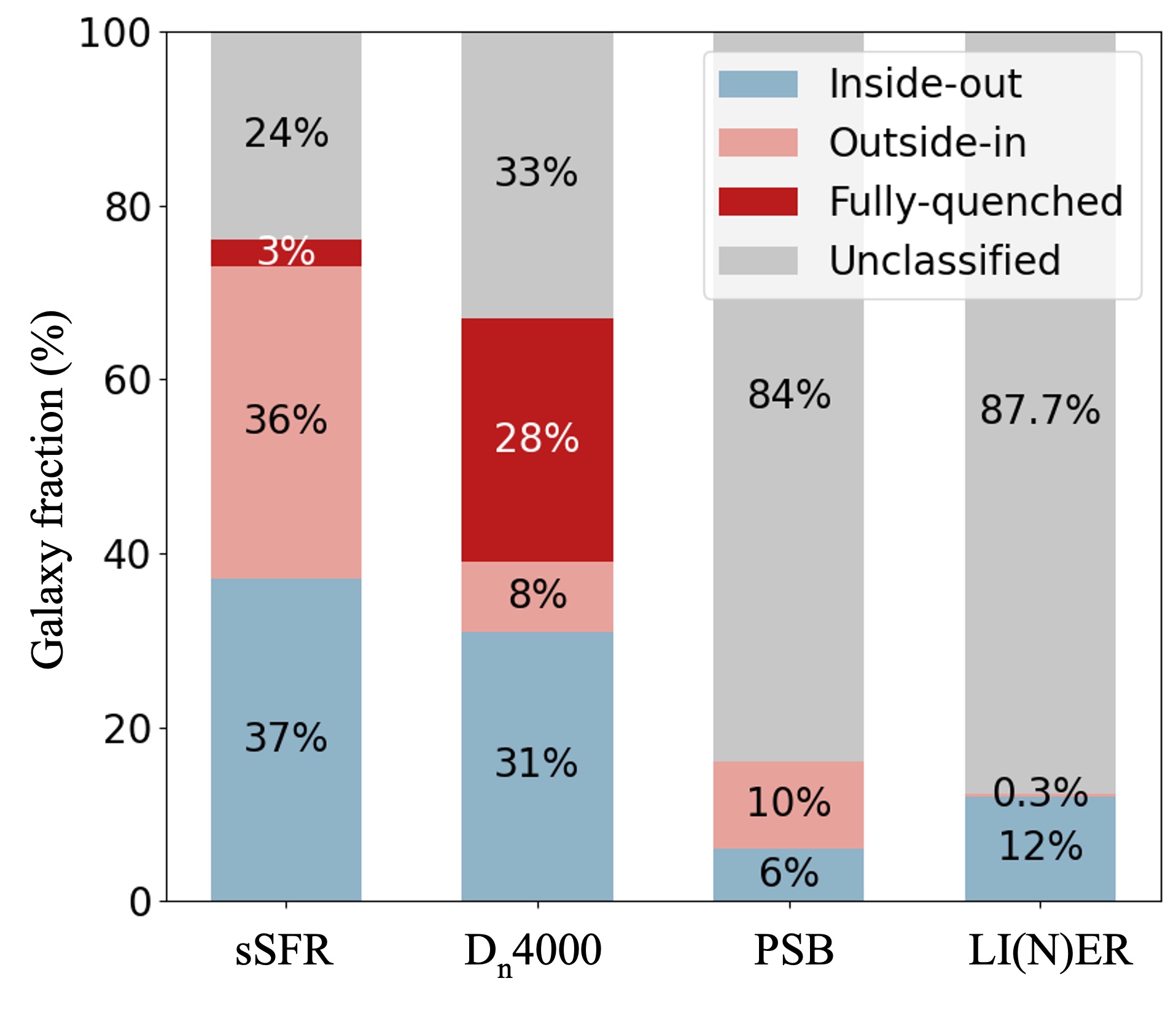}
    \caption{Fractions of galaxies classified into different spatial quenching modes under each criterion for the Full Sample. Blue bars represent inside-out quenching, red bars outside-in quenching, and dark red bars fully quenched galaxies. Galaxies with fewer than six quenched spaxels are labeled as “Unclassified” (gray).}
    \label{fig:barchart_full}
\end{figure}

We begin by presenting the spatial quenching classifications for the full MaNGA DR17 sample (10,220 galaxies), without imposing any overlap requirement among the different quenching criteria. This provides a global baseline against which the effects of stricter selection (e.g., Subsample A and Subsample B) can later be evaluated.

Figure~\ref{fig:fullsample} shows the distribution of galaxies in the $\mathrm{F}_q$–$\mathrm{C}_q$ plane for each of the four quenching criteria. The colors represent the average global sSFR of galaxies within each hexagonal bin, computed from the integrated measurements provided in the Pipe3D DR17 catalog. Figure~\ref{fig:barchart_full} summarizes the fraction of galaxies classified into inside-out, outside-in, fully quenched, or unclassified categories under each diagnostic. These fractions are computed at the galaxy level, based on the spatial quenching classification within each system. A galaxy is classified as fully quenched when all spaxels within 1.5~$R_e$ satisfy the corresponding quenching criterion, i.e., when $\mathrm{F}_q = 100\%$, where $\mathrm{F}_q$ denotes the fraction of quenched spaxels within a galaxy. Galaxies with fewer than six quenched spaxels under a given criterion are labeled as unclassified.

Substantial differences are observed across the four diagnostics in both the fraction of galaxies exhibiting significant quenched regions and the dominant spatial quenching mode. These variations highlight systematic differences among the diagnostics, which can be attributed to their distinct physical sensitivities and temporal responses. These differences are also reflected in the typical quenched area fractions inferred by each diagnostic. In the full sample, the sSFR and $D_n$4000 criteria generally extend to larger $\mathrm{F}_q$ values than the PSB and LI(N)ER criteria. This likely reflects the broader threshold-based definitions of sSFR and $D_n$4000, which trace star-formation suppression over comparatively extended phases, whereas the PSB and LI(N)ER criteria identify more restricted subsets of spaxels associated with a short-lived transitional phase or weak emission from evolved stellar populations.

From Figure~\ref{fig:fullsample} (top-left and top-right panels), galaxies selected by the sSFR and $D_n$4000 criteria span a broad range in global sSFR ($\log(\mathrm{sSFR}/\mathrm{yr}^{-1}) \sim -10.0$ to $-12.5$), indicating that these two diagnostics capture systems at various stages along the quenching sequence. 
In contrast, the PSB-selected galaxies (bottom-left panel) preferentially occupy the intermediate global sSFR regime, while LI(N)ER-selected galaxies (bottom-right panel) tend to reside at systematically lower global sSFR.

For the sSFR criterion (Figure~\ref{fig:barchart_full}), galaxies are nearly evenly divided between inside-out ($37\%\pm0.5\%$) and outside-in ($36\%\pm0.5\%$) quenching. Under the $D_n$4000 criterion, inside-out quenching is more common ($31\%\pm0.4\%$) than outside-in ($8\%\pm0.3\%$), and a significant fraction of galaxies are classified as fully quenched ($28\%\pm0.4\%$). 

Because the PSB criterion selects only a short-lived transitional phase rather than all quenched or quenching regions, it is not interpreted in the same way as the sSFR-, D$_n$4000-, or LI(N)ER-based classifications within the $F_q$–$C_q$ framework. Accordingly, differences between the PSB-based distribution and those derived from the other three diagnostics are expected. Instead, the PSB maps are used only to describe the spatial distribution of PSB-selected regions. Under this interpretation, galaxies with PSB-selected spaxels are more frequently located in the outside-in region of the $F_q$–$C_q$ plane ($10\%\pm0.3\%$) than in the inside-out region ($6\%\pm0.2\%$). This excess of PSB-selected galaxies in the outside-in region should therefore not be taken as a direct measurement of the underlying quenching sequence or dominant physical mechanism.

For the LI(N)ER criterion, inside-out quenching dominates ($12\%\pm0.3\%$), with a negligible outside-in component ($0.3\%\pm0.1\%$), qualitatively consistent with \citet{lin19}. Although the DR17 sample is approximately twice as large as the DR15 sample used in earlier work, the number of galaxies classified under the LI(N)ER criterion does not increase proportionally. This likely reflects the more conservative emission-line measurements in DR17, which are based on the MaStar stellar library with matched spectral resolution, reducing spurious detections at low signal-to-noise compared to earlier data releases \citep{san22}. As a result, the LI(N)ER-selected subsample remains relatively small.

\subsection{Quenching Mode Fractions in Subsample A}
\label{subsec:Quenching Mode Fractions in Subsample A}
\begin{figure*}[tbh]
    \centering
    \includegraphics[scale=0.162]{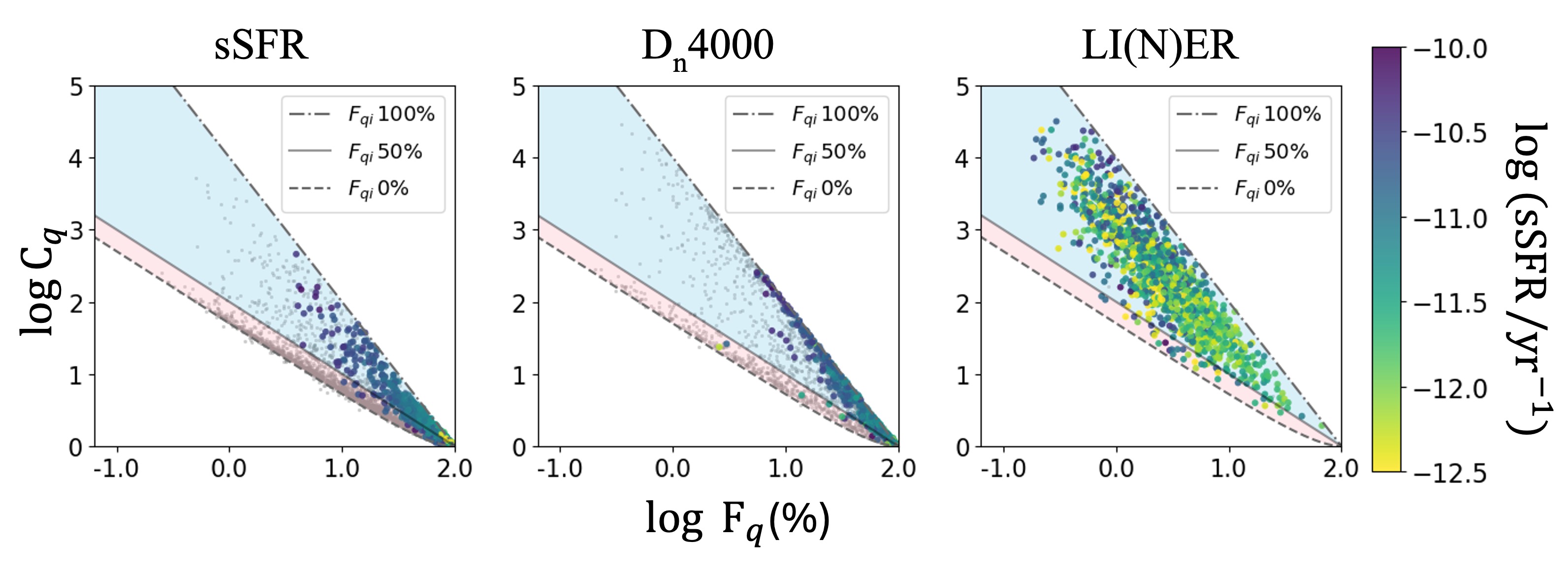}
    \caption{Distribution of galaxies in the $\mathrm{F}_q$--$\mathrm{C}_q$ plane for Subsample~A, shown separately for the three quenching criteria: sSFR (left), $D_n$4000 (middle), and LI(N)ER (right). Subsample~A consists of galaxies that simultaneously satisfy the sSFR, $D_n$4000, and LI(N)ER quenching criteria at the galaxy level (i.e., each contains at least six quenched spaxels under each diagnostic). The shaded regions indicate the inside-out (blue) and outside-in (pink) quenching regimes defined following \citet{lin19}. Gray points in the background indicate the corresponding full-sample distribution for each diagnostic. Each colored point represents one galaxy in Subsample~A and is color-coded by its global sSFR.}
    \label{fig:overlap3}
\end{figure*}

\begin{figure}[tbh]
    \centering
    \includegraphics[scale=0.127]{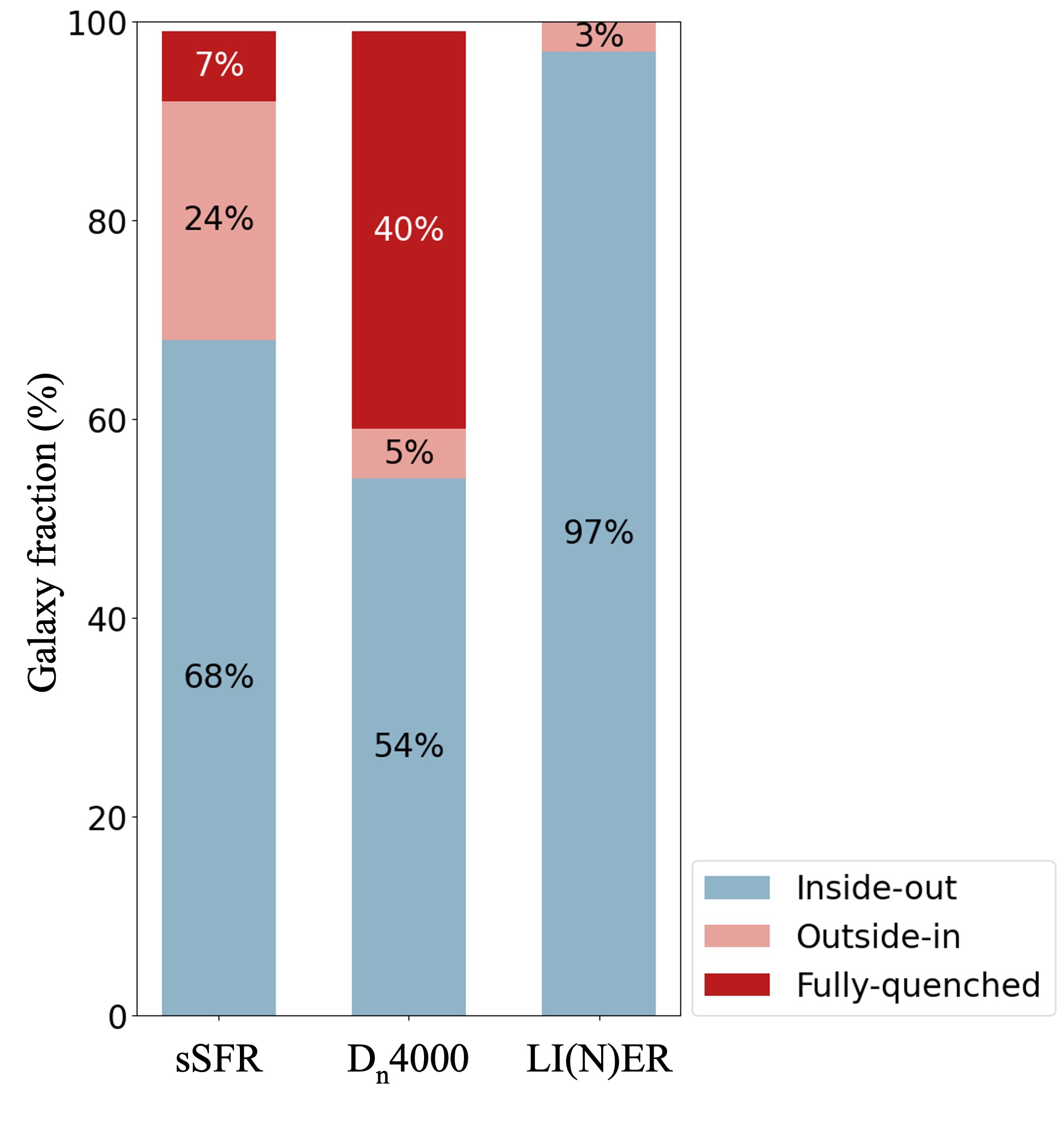}
    \caption{Galaxy fractions of spatial quenching classifications for Subsample~A under the three quenching criteria. Each bar is normalized by the total number of galaxies in Subsample~A and shows the fractions classified as inside-out (blue), outside-in (pink), fully quenched (red). Across all three diagnostics, Subsample~A is predominantly composed of inside-out quenched galaxies, with the outside-in fraction remaining subdominant.}
    \label{fig:barchart3}
\end{figure}

Subsample~A represents galaxies that exhibit significant quenched regions according to all three diagnostics that trace relatively long-lived quenching signatures (sSFR, $D_n$4000, and LI(N)ER), without requiring the presence of PSB features. As such, this sample serves as a baseline population of partially or fully quenched systems selected primarily by indicators sensitive to the overall suppression of star formation rather than short-lived transitional phases. A total of 1234 galaxies meet this stringent overlap requirement.

Figure~\ref{fig:overlap3} shows that galaxies in Subsample~A occupy systematically different regions of the $\mathrm{F}_q$--$\mathrm{C}_q$ plane depending on the diagnostic. Under the sSFR and $D_n$4000 criteria, most systems are concentrated at relatively high $\mathrm{F}_q$ values with a limited spread in $\mathrm{C}_q$, indicating that Subsample~A is dominated by galaxies that are already significantly quenched under these definitions. In contrast, the LI(N)ER criterion exhibits a broader distribution toward lower $\mathrm{F}_q$ values in the $\mathrm{F}_q$--$\mathrm{C}_q$ plane, indicating that many systems in Subsample~A are less fully quenched under this diagnostic.

This trend is quantified in Figure~\ref{fig:barchart3}. For the sSFR criterion, inside-out quenching is more common than outside-in quenching (68.23$\pm$1.3\% vs.\ 24.15$\pm$1.2\%), with a smaller fraction of fully quenched systems (7.62$\pm$0.8\%), indicating a relatively balanced distribution of quenching modes. The $D_n$4000 criterion shows a comparable fraction of inside-out quenching (54.29$\pm$1.4\%), but a substantially larger fraction of fully quenched systems (40.44$\pm$1.4\%), while outside-in quenching becomes relatively rare (5.27$\pm$0.6\%). 

The LI(N)ER criterion is overwhelmingly dominated by inside-out classifications (97.33$\pm$0.5\%), with only a very small fraction of outside-in systems (2.67$\pm$0.5\%) and no galaxies classified as fully quenched. This indicates that the LI(N)ER selection strongly favors centrally concentrated quenching structures within Subsample~A, rather than providing a balanced sampling of different quenching modes.

\subsection{Quenching Mode Fractions in Subsample B}
\label{subsec:Quenching Mode Fractions in Subsample B}
\begin{figure*}[tbh]
    \centering
    \includegraphics[scale=0.205]{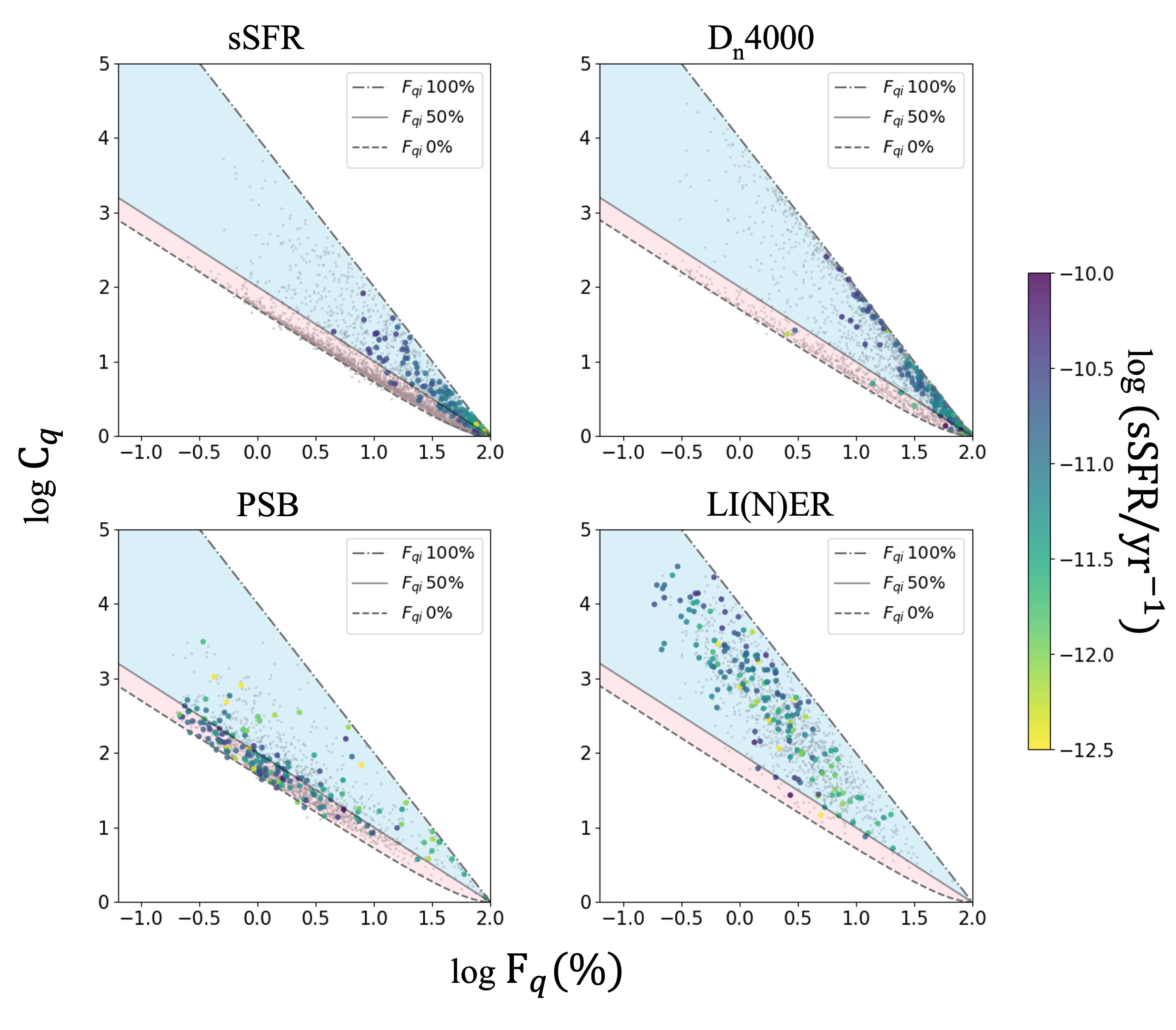}
    \caption{$\mathrm{F}_q$--$\mathrm{C}_q$ plane color-coded by $\log(\mathrm{sSFR}/\mathrm{yr}^{-1})$ of Subsample~B (from left to right, top to bottom: sSFR, D$_n$4000, PSB, and LI(N)ER). The x-axis represents $\mathrm{F}_q$, the fraction of selected area, where a higher $\mathrm{F}_q$ indicates a larger area identified by the corresponding criterion. The y-axis represents $\mathrm{C}_q$, the concentration of the selected area, with lower $\mathrm{C}_q$ indicating a more centrally concentrated spatial distribution. Gray points in the background indicate the corresponding full-sample distribution for each diagnostic. The corresponding galaxy fractions are shown in Figure~\ref{fig:barchart}.}
    \label{fig:overlap}
\end{figure*}

\begin{figure}[tbh]
    \centering
    \includegraphics[scale=0.125]{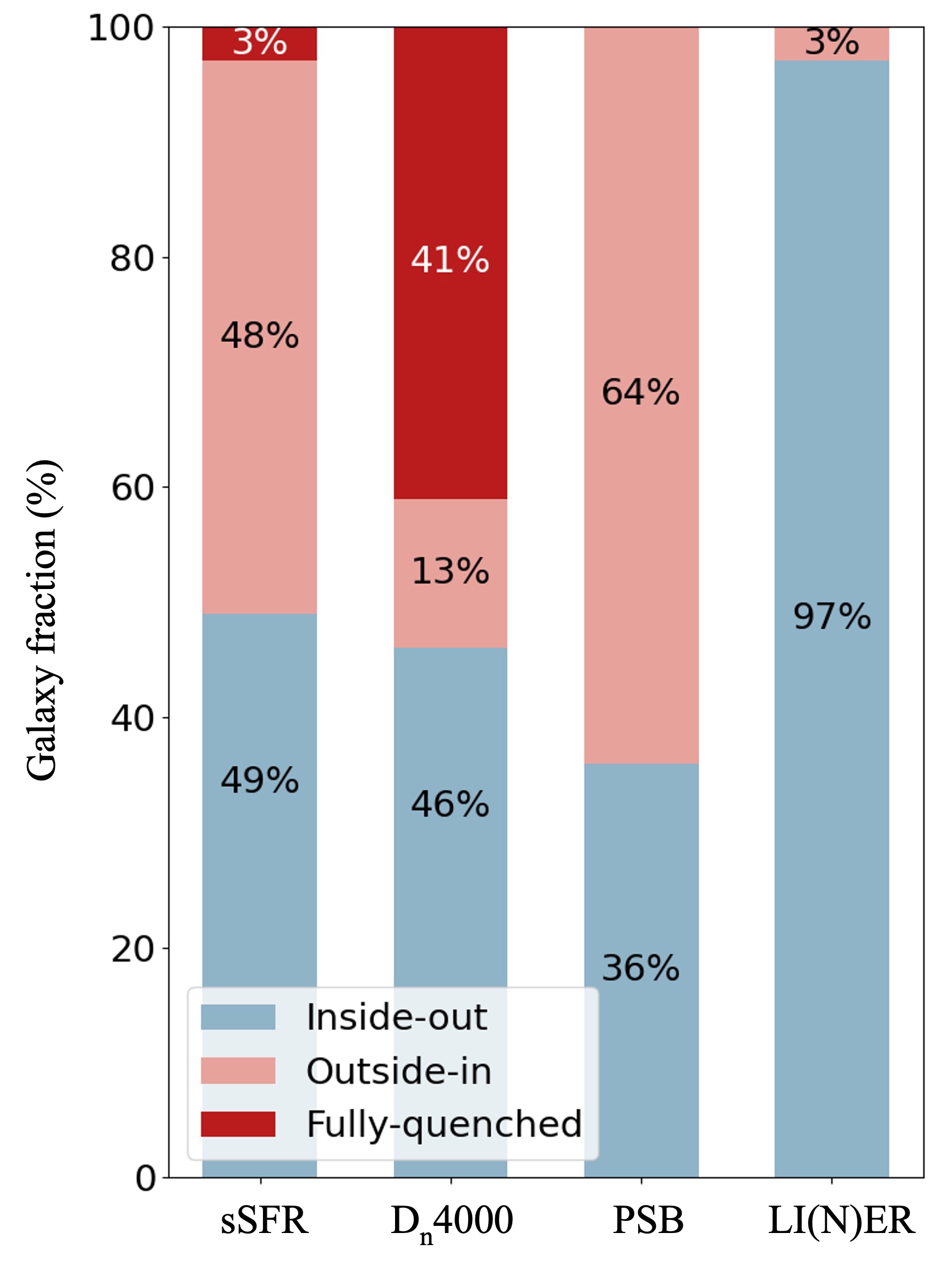}
    \caption{Quenching mode fractions for galaxies that satisfy all criteria. The y-axis indicates the fraction of galaxies exhibiting each quenching mode, normalized by the total number of overlapping galaxies (172 galaxies). Blue bars represent inside-out quenching, red bars correspond to outside-in quenching, and dark red bars are fully quenched galaxies. For the sSFR criterion, the inside-out and outside-in quenching mode fractions are comparable. As for the D$_n$4000 and LI(N)ER criteria, the majority of galaxies are classified as inside-out quenched. In contrast, the PSB criterion yields a higher fraction of outside-in quenching galaxies.}
    \label{fig:barchart}
\end{figure}

Subsample~B consists of galaxies that satisfy all four quenching criteria simultaneously (sSFR, $D_n$4000, PSB, and LI(N)ER), representing the most restrictive selection in this work. This requires galaxies to simultaneously satisfy diagnostics sensitive to different stages of quenching: recent or ongoing star-formation suppression traced by sSFR, cumulative stellar population aging traced by $D_n$4000, short-timescale transitional features traced by PSB, and LI(N)ER-like emission associated with evolved stellar populations. Because PSB-selected galaxies in our sample preferentially occupy the intermediate global sSFR regime (Figure~\ref{fig:overlap}), this selection naturally biases Subsample~B toward green-valley systems. A total of 172 galaxies meet this stringent overlap requirement.

Figure~\ref{fig:overlap} presents the distribution of Subsample~B galaxies in the $\mathrm{F}_q$--$\mathrm{C}_q$ plane under each criterion. Compared to the full galaxy sample, Subsample~B occupies a more restricted region of the $\mathrm{F}_q$–$\mathrm{C}_q$ plane, reflecting the combined constraints imposed by all four diagnostics. In particular, galaxies with very low $\mathrm{F}_q$ values are less common in this subsample. We also note that the overall distribution of Subsample~B is broadly similar to that of Subsample~A, indicating that the additional PSB requirement does not substantially shift the galaxies to a different region of the $\mathrm{F}_q$--$\mathrm{C}_q$ plane, but mainly further reduces the sample size.

The corresponding quenching mode fractions for Subsample~B are summarized in Figure~\ref{fig:barchart}. Under the sSFR criterion, inside-out and outside-in quenching occur at comparable frequencies (55.48$\pm$0.6\% vs.\ 40.48$\pm$0.6\%), with a small fraction of fully quenched systems (4.04$\pm$0.2\%). The $D_n$4000 criterion shows a comparable fraction of inside-out quenching (44.95$\pm$0.6\%), but a substantially larger fraction of fully quenched systems (43.01$\pm$0.6\%), while outside-in quenching is less common (12.04$\pm$0.4\%).

In contrast, the PSB criterion is dominated by outside-in classifications (63.57$\pm$1.6\%), with a smaller fraction of inside-out systems (36.43$\pm$1.6\%) and no galaxies classified as fully quenched. The LI(N)ER criterion shows the opposite behavior, being overwhelmingly dominated by inside-out quenching (97.33$\pm$0.5\%), with only a very small fraction of outside-in systems (2.67$\pm$0.5\%). These results are also consistent with the broadly similar distributions of Subsamples~A and B in the $\mathrm{F}_q$--$\mathrm{C}_q$ plane.

\color{black}
In terms of quenched area fraction, $\mathrm{F}_q$ values derived from the sSFR and $D_n$4000 indicators are generally larger than those from the PSB and LI(N)ER criteria. This difference arises primarily from the restrictive nature of the multi-diagnostic selection and the specific sensitivities of the individual diagnostics. Systems with only a small fraction of quenched regions are less likely to satisfy all four criteria simultaneously, naturally reducing the presence of low-$\mathrm{F}_q$ galaxies in Subsample~B. This selection effect acts on top of the more general tendency, already seen in the full sample, for the sSFR and $D_n$4000 criteria to yield larger $\mathrm{F}_q$ values than the PSB and LI(N)ER criteria.

\subsection{The relation between quenching pattern, halo mass, and stellar mass}
\label{subsec:The relation between quenching pattern, halo mass, and stellar mass}
\begin{figure*}[tbh]
    \centering
    \includegraphics[scale=0.22]{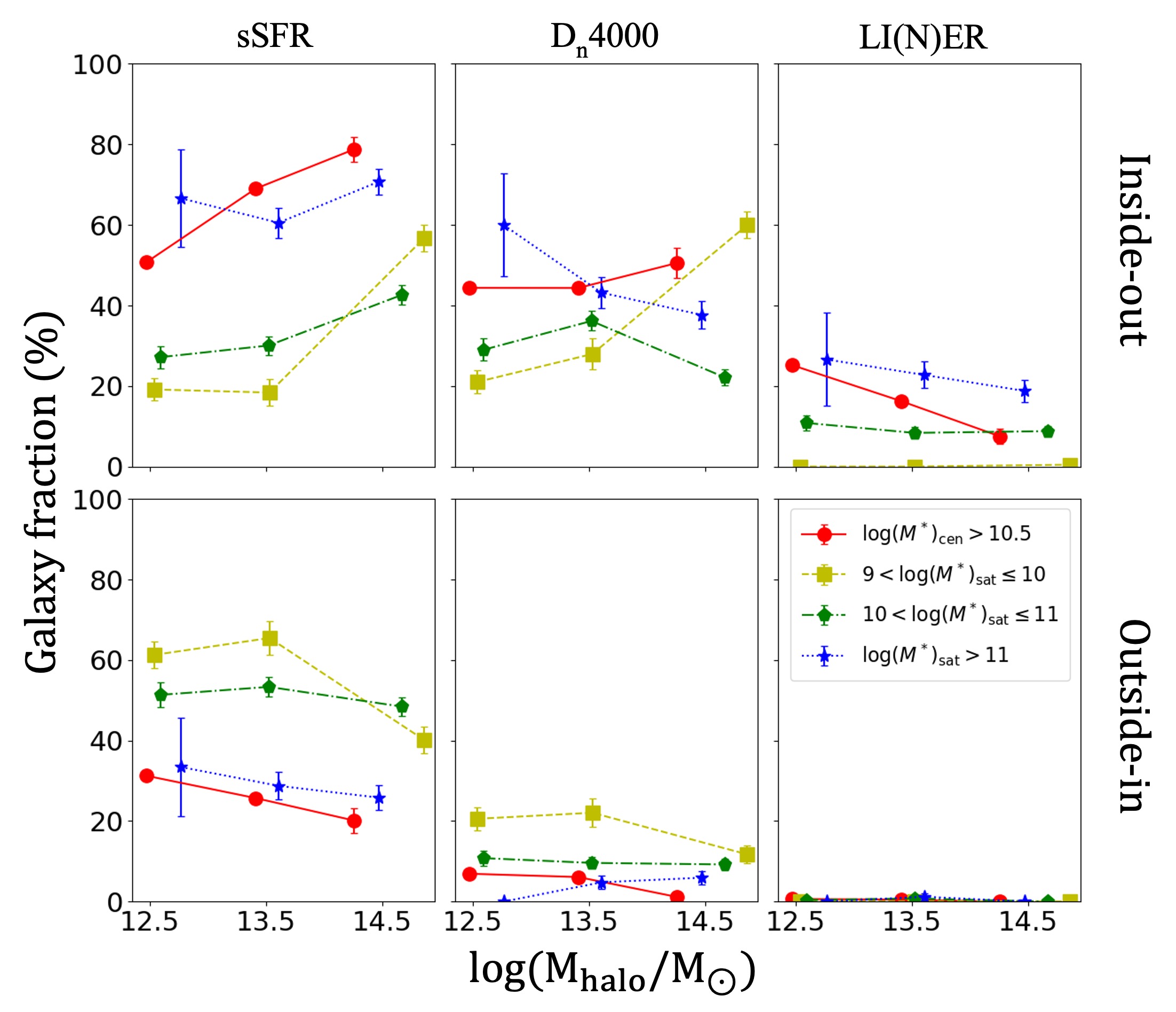}
    \caption{Fractions of galaxies classified as inside-out (top row) and outside-in (bottom row) as a function of halo mass for three quenching criteria (from left to right: sSFR, $D_n$4000,  and LI(N)ER). Colored lines represent central galaxies $\log(M)_{\rm cen} > 10.5$ (red) and satellite galaxies in different stellar-mass bins: $9 < \log(M)_{\rm sat} \leq 10$ (yellow), $10 < \log(M)_{\rm sat} \leq 11$ (green), and $\log(M)_{\rm sat} > 11$ (blue).}
    \label{fig:hmass}
\end{figure*}

We now examine how the fraction of galaxies classified as inside-out or outside-in depends on halo mass and stellar mass under each quenching criterion (Figure~\ref{fig:hmass}). 

The diagnostics considered in this work probe different stages of star-formation suppression. While sSFR and $D_n$4000 primarily identify galaxies with already suppressed star formation, and LI(N)ER emission is generally associated with more evolved, quiescent systems, the PSB criterion selects galaxies undergoing or having very recently experienced rapid quenching. 

The PSB diagnostic traces a fundamentally different evolutionary phase compared to the other three criteria. As a result, the halo mass and stellar mass dependence derived from the PSB criterion should be interpreted as tracing transitional systems, and is therefore not included in Figure~\ref{fig:hmass}. In contrast, the relations derived from the other three diagnostics primarily reflect more established quenched populations.

\medskip
\noindent \textbf{sSFR criterion.}
Under the sSFR criterion (left column of Figure~\ref{fig:hmass}), the fraction of galaxies classified as inside-out generally increases with halo mass for both centrals (red line, top-left panel) and massive satellites (blue line, top-left panel). In contrast, the outside-in fraction (bottom-left panel) decreases with halo mass for centrals and high-mass satellites. Low-mass satellites (yellow line) show a non-monotonic trend, with outside-in fractions peaking at intermediate halo mass.

In addition, a clear stellar-mass dependence is present: at fixed halo mass, the inside-out fraction increases with stellar mass (compare yellow, green, and blue lines in the top-left panel), while the outside-in fraction decreases correspondingly (bottom-left panel).

\medskip
\noindent \textbf{$D_n$4000 criterion.}
For the $D_n$4000 criterion (middle column), a similar stellar-mass dependence is observed: massive galaxies preferentially exhibit inside-out quenching (top-middle panel), while lower-mass galaxies show higher outside-in fractions (bottom-middle panel). However, unlike the sSFR case, no clear monotonic trend with halo mass is seen across most stellar-mass bins.

\medskip
\noindent \textbf{LI(N)ER criterion.}
For the LI(N)ER criterion (right column), inside-out quenching clearly dominates across all stellar-mass bins (top-right panel). The inside-out fraction increases with stellar mass, consistent with the trends seen for sSFR and $D_n$4000. However, for central galaxies (red line), the inside-out fraction decreases with increasing halo mass (top-right panel), in contrast to the sSFR-based trend. This behavior differs from \citet{lin19} and is likely related to systematic differences between the data products (DR15 vs. DR17) used in the two studies \citep{san22}.

Overall, stellar mass exhibits a stronger and more consistent correlation with quenching pattern than halo mass across all three diagnostics. Halo-mass trends, when present, are generally secondary and more sensitive to the choice of quenching criterion.


\section{discussion}
\label{sec:discussion}

\subsection{Quenching signatures under different criteria}
\label{subsec:Quenching signatures under different criteria}
\color{black}
We apply four different criteria (sSFR, $D_n$4000, PSB, and LI(N)ER) to classify spaxels under different quenching stages and investigate how different diagnostics identify quenching within MaNGA galaxies. In this section, we focus on the differences in sample size, global sSFR distribution, and quenching mode among the four diagnostics. The dependence on stellar and halo mass will be discussed in \S\ref{subsec:Stellar mass and halo mass dependence}.

\subsubsection{The number of selected galaxies}
\label{subsubsec:The number of selected galaxies}

The variation in the number of galaxies identified under each criterion likely reflects the different timescales and physical sensitivities of each indicator. The sSFR criterion ($N = 7766$) measures the current SFR relative to the stellar mass and therefore responds to changes in ongoing star formation activity. As shown in Appendix~\ref{app:sfh_timescales}, the sSFR threshold is sensitive to the rate of star formation decline, with more rapidly quenched systems reaching the threshold at earlier times. However, because the sSFR criterion is defined by a threshold on the current star formation level, galaxies undergoing a wide range of quenching histories—whether rapid or gradual—will eventually satisfy this condition once their star formation is sufficiently suppressed. As a result, the sSFR selection includes systems at different stages of quenching, from galaxies that have only recently begun to decline in star formation to those that have already reached more quiescent states. This broad inclusiveness naturally leads to the largest sample size among all indicators.

The D$_n$4000 index ($N = 6826$) is an age-sensitive stellar population indicator that increases as stellar populations age, and is therefore more likely to select regions dominated by intermediate-age to old stars under our adopted threshold \citep{pog97,kau03}. Compared with sSFR, the $D_n$4000 diagnostic responds differently to changes in the recent star formation history. In our toy SFH models, the $D_n$4000 threshold is often reached earlier than the sSFR threshold under the adopted cuts. However, the sSFR evolution changes more strongly when the quenching timescale is varied, whereas $D_n$4000 evolves more gradually. The difference between the two diagnostics therefore lies not simply in when their thresholds are crossed, but in how their responses map onto the recent quenching history and, in turn, onto the spatially resolved galaxy classification.

PSB features ($N = 884$), characterized by strong Balmer absorption lines from A-type stars, trace recent starbursts followed by rapid quenching within the last $\sim$1 Gyr \citep{cou87, dre83, pog99}. This phase is both short-lived and requires a specific evolutionary pathway (i.e., a recent burst followed by rapid suppression of star formation), making PSB galaxies intrinsically rare and leading to a significantly smaller sample size (see Appendix~\ref{app:sfh_timescales}).

LI(N)ER features ($N = 1234$) (after excluding AGN contamination) are generally associated with ionization by hot evolved stars such as post-AGB stars, and are therefore expected to appear in late evolutionary stages after star formation has largely ceased \citep{bel16, mil16, byl19}. Although this phase can persist over relatively long timescales, it only occurs once galaxies are already fully quenched, making it more restrictive than sSFR and D$_n$4000 in selecting quenching systems.

In summary, these differences can be understood in terms of an evolutionary sequence and corresponding timescales: sSFR traces the earliest and broadest range of quenching stages, D$_n$4000 selects more evolved populations, PSB isolates a brief transitional phase, and LI(N)ER corresponds to the latest stages of quenching. This naturally explains the observed ranking of sample sizes.

\subsubsection{The global sSFR distribution}
\label{subsubsec:The global sSFR distribution}
Galaxies selected by the sSFR and D$_n$4000 criteria exhibit a broad distribution of global sSFR, encompassing star-forming, green valley, and quiescent populations, as shown in Figure~\ref{fig:fullsample}. In contrast, PSB-selected galaxies are concentrated at relatively high global sSFR, while LI(N)ER-selected galaxies are skewed toward lower global sSFR. This suggests that PSB galaxies are not fully quenched, consistent with the idea that PSB signatures can arise in galaxies that are only partially quenched or have just undergone quenching recently \citep{chen19,cheng24}.

This result differs from early studies based on single-fiber spectroscopy \citep[e.g.,][]{goto03, won12}, which typically interpreted PSB galaxies as systems with little or no ongoing star formation and as transitioning toward the quiescent population. In our case, however, galaxies are selected from IFU data and are only required to contain at least six PSB-selected spaxels within 1.5~$R_e$, rather than having their integrated spectra satisfy the PSB criterion. This means that our sample can include galaxies with only a localized PSB signature, and is therefore broader than the samples identified in earlier single-fiber studies. However, our finding is consistent with spatially resolved studies using IFU spectroscopy \citep[e.g.,][]{chen19, cheng24}, which reveal that IFU-selected PSB galaxies frequently lie on the star-forming main sequence or in the green valley, and represent transitional systems.

\subsubsection{The dominant quenching mode}
\label{subsubsec:The dominant quenching mode}

The dominant quenching mode varies with the selection criterion, reflecting the sensitivity of each tracer to different quenching mechanisms and timescales.

For galaxies classified using the sSFR criterion, the fractions of inside-out and outside-in quenching are comparable. Inside-out quenched galaxies exhibit a smooth decline in global sSFR with increasing $F_q$, forming a continuous sequence from partially to nearly fully quenched systems.

This difference likely reflects the characteristic timescales of the underlying quenching processes. Inside-out quenching is often interpreted as being associated with gradual, internally driven mechanisms, such as morphological quenching or AGN feedback \citep[e.g.,][]{mar09, fab12, blu16}. By contrast, outside-in quenching is commonly linked to environmental processes, such as ram-pressure stripping or gas removal \citep[e.g.,][]{gun72, pen12}, which can act on shorter timescales. As a result, galaxies undergoing outside-in quenching may transition rapidly from star-forming to fully quenched states, spending relatively little time at intermediate sSFR, leading to a deficit of low-sSFR systems classified as outside-in in our sample.

For galaxies classified using the D$_n$4000 criterion, the inside-out fraction remains dominant, while the fully quenched fraction increases substantially and the outside-in fraction is reduced compared to the sSFR classification. A transition analysis shows that the D$_n$4000 fully quenched sample is still composed predominantly of galaxies classified as inside-out under the sSFR criterion mainly because the sSFR inside-out population is larger than the sSFR outside-in population in the matched sample (3683 versus 2687 galaxies), while the probabilities of being classified as fully quenched under the D$_n$4000 criterion are very similar for the two classes (42.5\% for sSFR inside-out and 40.5\% for sSFR outside-in). Given the strong overlap between the sSFR and D$_n$4000 selections at the spaxel level (Figure~\ref{fig:upsetplot}), this indicates that the two diagnostics identify similar quenched regions but do not map identical galaxy populations into the same late-stage quenching classes. The higher fully quenched fraction under the D$_n$4000 criterion is broadly consistent with the toy SFH models in Appendix~\ref{app:sfh_timescales}, which show that the adopted D$_n$4000 threshold is often reached earlier than the sSFR threshold. This suggests that galaxies can be classified as fully quenched under D$_n$4000 even when residual spatial structure is still present in the sSFR map. By contrast, we do not assign a unique physical interpretation to the reduced outside-in fraction under the D$_n$4000 criterion, and instead describe it more conservatively as an empirical difference between the two classification schemes.

Among D$_n$4000-selected inside-out galaxies, systems with lower global sSFR tend to have higher $F_q$, forming a sequence similar to that seen in the sSFR classification. Because D$_n$4000 is sensitive to the mean stellar age and star formation over the past few Gyr \citep{bal99,kau03,wu18}, rather than instantaneous star formation, it more readily classifies galaxies with broadly old stellar populations as fully quenched, even when the sSFR map still retains inside-out or outside-in structure. By contrast, the smaller outside-in population selected by D$_n$4000 tends to occupy higher global sSFR and lower $F_q$. We interpret this more conservatively as indicating that the sSFR and D$_n$4000 criteria do not map identical late-stage quenching populations, rather than assigning it a unique physical origin.

PSB classification is also sensitive to localized star formation or recent starburst activity, which can dominate spaxel-level spectral features. As a result, it preferentially traces stochastic or spatially heterogeneous processes rather than a coherent, galaxy-wide quenching pattern. The high fraction of unclassified systems is therefore primarily driven by the stringent selection criteria and the short-lived nature of the PSB phase. This interpretation is also consistent with resolved PSB studies that find a large fraction of morphologically irregular PSB systems \citep{chen19, cheng24}. We thus treat PSB as a tracer of rapid quenching events rather than a primary quenching diagnostic.

We note that the inside-out/outside-in classes in our PSB analysis are defined from the non-parametric $F_q$--$C_q$ framework and are not directly equivalent to the central/ring-like/irregular PSB morphological classes used in previous resolved PSB studies. Because our method does not require the PSB-selected spaxels to be contiguous, ring-like PSB configurations can appear in both the inside-out and outside-in regions of the $F_q$--$C_q$ plane. We further find that many PSB-selected outside-in galaxies are classified as inside-out under the $D_n$4000 criterion. This is qualitatively consistent with the idea that, in some galaxies, central regions may already have evolved beyond the PSB-selection window while outer regions still retain PSB signatures. However, because the correspondence between the PSB-based and sSFR/$D_n$4000-based classifications is not one-to-one, we do not regard this as a unique explanation for the full sample.

Galaxies selected by the LI(N)ER criterion exhibit a strong preference for inside-out quenching, with minimal outside-in signatures. These galaxies typically have low global sSFR and moderate $\mathrm{F}_q$, consistent with retired stellar populations and a late stage of quenching \citep{yan12, cid11, bel16}. To better understand the strong inside-out preference of the LI(N)ER criterion, we examined the radial distribution of LI(N)ER-selected spaxels and found that they are already strongly concentrated toward the inner regions. We also compared the all-valid and BPT-valid parent samples, and found that the BPT-valid sample is somewhat more centrally concentrated than the all-valid sample, indicating a modest radial bias in emission-line measurability. However, the normalized radial profiles of LI(N)ER-selected spaxels remain very similar when evaluated with respect to these two parent samples. This suggests that the high inside-out fraction of the LI(N)ER selection is unlikely to be driven primarily by the radial dependence of emission-line detectability alone, and is instead more consistent with the intrinsically central concentration of LI(N)ER-like emission in these galaxies, although the present analysis does not uniquely determine which physical ingredient is dominant. Compared to the sSFR and D$_n$4000 criteria, LI(N)ER-selected systems therefore represent more evolved quiescent populations.

\subsection{Stellar mass and halo mass dependence}
\label{subsec:Stellar mass and halo mass dependence}
As qualitatively discussed in \S\ref{subsec:The relation between quenching pattern, halo mass, and stellar mass}, the quenching behavior of galaxies selected under different criteria shows different trends with stellar mass and halo mass. In this section, we summarize these trends in a more structured way, discuss their possible physical origins, and compare our results with previous studies.

\subsubsection{Stellar mass dependence}
\label{subsubsec:Stellar mass dependence}

Among the sSFR, D$_n$4000, and LI(N)ER criteria, the stellar-mass dependence of the quenching mode fraction appears broadly consistent for satellite galaxies: the inside-out quenching mode fraction increases with stellar mass at fixed halo mass. For central galaxies, however, our sample contains only a single stellar-mass bin, and therefore does not allow us to robustly assess any stellar-mass dependence. 

The trend observed for satellites is broadly consistent with the commonly accepted picture in which mass-related quenching processes—such as AGN feedback and morphological quenching—become increasingly important in more massive galaxies \citep{con20,blu20,mao22}, in agreement with earlier work \citep{lin19}.

Conversely, outside-in quenching in the sSFR and D$_n$4000 criteria exhibits the opposite trend: its fraction decreases with increasing stellar mass at fixed halo mass. This behavior is largely driven by the satellite population, where environmental mechanisms—such as ram pressure stripping or tidal interactions—are more effective in low-mass systems, often initiating quenching from the outskirts. Notably, high-mass satellites display trends more similar to central galaxies, a behavior also noted by \citet{blu20}. In the LI(N)ER criterion, the scarcity of outside-in quenched galaxies makes it difficult to interpret any mass-dependent trends reliably.

\subsubsection{Halo mass dependence}
\label{subsubsec:Halo mass dependence}

After examining the stellar-mass dependence of quenching modes, we next investigate how these trends vary with halo mass. As shown in Figure~\ref{fig:hmass}, the halo-mass dependence of inside-out and outside-in quenching mode fractions differs substantially between the sSFR and D$_n$4000 criteria.

In particular, the sSFR-based classification shows a clearer increase in the inside-out quenching fraction with halo mass, whereas the D$_n$4000-based results display a weaker or absent trend. This difference likely reflects the distinct physical sensitivities of the two diagnostics. The difference between the sSFR and D$_n$4000 criteria is better understood in terms of how they respond to changes in recent star-formation history, rather than simply in terms of which threshold is crossed first. In our toy SFH models, the adopted D$_n$4000 threshold is often reached earlier than the sSFR threshold. However, when the quenching timescale is varied, the sSFR evolution changes more strongly, whereas the D$_n$4000 evolution is more gradual because it reflects the cumulative aging of the stellar population. In this sense, sSFR is more sensitive to differences in recent quenching timescale, while D$_n$4000 provides a smoother record of the integrated recent star-formation history.

We further compare our results with those of \citet{lin19}, who used LI(N)ER emission as the primary quenching diagnostic based on the DR15 Pipe3D data products. Their results show a stronger and more systematic increase of inside-out quenching with halo mass, particularly for central galaxies. In contrast, our DR17-based analysis yields a weaker trend in the LI(N)ER-selected sample.

This discrepancy is likely driven by differences in emission-line measurements between DR15 and DR17. The DR17 Pipe3D products adopt an updated SSP library based on the \texttt{MaStar\_sLOG} library, which better matches the spectral resolution of MaNGA data and improves the modeling of low signal-to-noise spectra. This leads to more conservative measurements of weak emission lines such as H$\alpha$, reducing spurious or marginal detections that can otherwise be classified as LI(N)ER-like emission \citep{san22}. As a result, LI(N)ER classifications in DR17 are generally more stringent, leading to fewer identified LI(N)ER spaxels and a reduced inside-out quenching fraction compared to DR15-based results. Because both the LI(N)ER highly rely on emission-line measurements, it is more sensitive to these data-processing differences, whereas the D$_n$4000 diagnostic, which is based on stellar continuum properties, is largely unaffected. 

To summarize, these results highlight that the inferred environmental dependence of quenching modes depends not only on the choice of diagnostic but also on the robustness of emission-line measurements. Care must therefore be taken when interpreting halo-mass trends, as both tracer sensitivity and data-processing effects can introduce systematic differences.

\subsection{From Full Sample to Subsample}
\label{subsec:From full to overlapping sample}
We present a comparative analysis of overlapping galaxies—those that simultaneously satisfy all four quenching criteria (sSFR, D$_n$4000, PSB, and LI(N)ER)—and the full MaNGA sample. By examining their quenching modes, global properties, and spatial characteristics, we explore the biases and insights that emerge when using multiple tracers in combination.

\subsubsection{Impact of multi-tracer selection on sample properties}

We now examine how the galaxy population changes when progressively applying multiple quenching diagnostics, from the full sample (satisfying at least one criterion) to subsample A (sSFR $\cap$ D$_n$4000 $\cap$ LI(N)ER) and subsample B (sSFR $\cap$ D$_n$4000 $\cap$ PSB $\cap$ LI(N)ER).

Compared to the full sample, the multi-tracer selections lead to a substantial reduction in sample size, reflecting the increasingly restrictive requirement that galaxies simultaneously satisfy multiple quenching criteria (see \S\ref{subsec:Quenching Mode Fractions in Subsample A} and \S\ref{subsec:Quenching Mode Fractions in Subsample B}). Along with this tightening of selection, the population properties also shift systematically. In particular, the fraction of fully quenched galaxies increases significantly in the D$_n$4000-based classification. This increase is mainly a consequence of the overlap selection itself. In the full sample, a substantial fraction of galaxies are unclassified under the D$_n$4000 criterion because they contain fewer than six selected spaxels. Once the overlap requirement is imposed, these galaxies are preferentially removed, so the remaining sample is biased toward systems with larger quenched areas and more globally evolved stellar populations. As a result, the fully quenched fraction under the D$_n$4000 criterion naturally increases in Subsamples A and B. This interpretation is also consistent with the shift toward higher $F_q$ values seen in Figures \ref{fig:overlap3} and \ref{fig:overlap}.

At the same time, the quenching mode distribution becomes less balanced compared to the full sample. While the full sample includes a mixture of inside-out and outside-in quenching systems (see Figure~\ref{fig:fullsample}), subsample A and B increasingly favor centrally quenched, inside-out configurations. This shift can be attributed in part to the LI(N)ER selection, which preferentially traces galaxies with centrally concentrated old stellar populations. Although the inclusion of PSB in subsample B introduces sensitivity to more recent quenching events, its impact remains limited due to its shorter characteristic timescale and more localized spatial signatures.

These trends are further illustrated by the spatially resolved maps of three representative galaxies from subsample B (Figure~\ref{fig:maps}). While all selected galaxies satisfy the multi-tracer criteria, the spatial distributions identified by different diagnostics are not identical within individual systems. The sSFR and D$_n$4000 selections generally trace more extended and contiguous quenched regions. The LI(N)ER classification, however, shows a strong preference for inside-out quenching patterns, as indicated by the F$_q$–C$_q$ analysis and the quenching mode distributions (Figure~\ref{fig:barchart_full}). 

We emphasize that these examples are intended to illustrate the diversity of spatial patterns within the overlapping sample, rather than to define a universal configuration. The coexistence of different spatial signatures within the same galaxy highlights that the diagnostics probe distinct physical conditions, even when they select the same system at the global level.

Overall, the multi-tracer selection does not simply refine the same galaxy population identified by individual diagnostics. Instead, it preferentially selects galaxies that simultaneously satisfy criteria probing different phases of quenching. In this sense, the transition from the full sample to Subsamples~A and B can be understood as an intersection of broader tracers such as sSFR and D$_n$4000 with the much more restrictive PSB selection window. As a result, Subsample~B is better interpreted as a population biased toward galaxies that still show relatively recent PSB signatures, while simultaneously lying within the broader sSFR and D$_n$4000 selection windows.

\subsubsection{Advantages and limitations of multi-tracer selection}

The impact of a multi-tracer selection depends strongly on the level at which it is interpreted. At the level of individual galaxies, requiring consistency across multiple diagnostics provides a clear advantage. It reduces the likelihood of misclassification caused by contamination or measurement uncertainties in any single tracer, and therefore yields a subset of systems that can be more confidently identified as quenched.

However, at the population level, this same requirement introduces a systematic selection bias. Rather than sampling the full diversity of quenching pathways, the overlapping sample preferentially selects galaxies that simultaneously satisfy multiple, and often heterogeneous, diagnostic conditions. As a result, the inferred properties of the sample do not simply reflect the underlying galaxy population, but are shaped by the combined sensitivities of the tracers.

In particular, diagnostics such as D$_n$4000 and LI(N)ER, which are sensitive to longer-term stellar evolution and to specific emission properties, tend to favor systems with extended quenching histories and inside-out quenching patterns. Meanwhile, diagnostics that rely on emission-line measurements (e.g., sSFR and LI(N)ER) are more susceptible to low signal-to-noise or non-detections in fully quenched regions, which can limit the identification of quenched spaxels in the F$_q$–C$_q$ framework. Conversely, short-timescale or spatially localized quenching signatures, such as those traced by PSB features, are less likely to be retained unless they coexist with longer-timescale indicators.

Therefore, the multi-tracer selection does not simply produce a cleaner version of the quenched galaxy population. Instead, it selects a more restricted subset of galaxies that simultaneously satisfy multiple diagnostic conditions, and is therefore biased toward specific quenching properties.

The main value of the multi-tracer approach lies not in the overlapping sample itself, but in the comparison between different diagnostics. As shown throughout this work, different tracers often identify different quenching modes, timescales, and spatial patterns. These discrepancies provide important clues to the underlying physical processes, such as the roles of emission-line dependence and the timescale sensitivity of each diagnostic.

By considering both the agreement and disagreement between tracers, we are able to better distinguish between different quenching pathways. In this context, the overlapping sample should be interpreted as a selected subset of galaxies that satisfy multiple conditions, rather than as a representative population of all quenched systems.


\color{black}
\section{conclusion}
\label{sec:conclusion}
In this study, we investigated how different definitions of quenching affect the interpretation of spatially resolved quenching modes in galaxies, using a sample of 10,220 galaxies from MaNGA DR17. By applying four widely used quenching diagnostics—sSFR, D$_n$4000, PSB, and LI(N)ER—we analyzed the spatial distribution of quenched regions through the $F_q$ and $C_q$ parameters, where $F_q$ represents the fraction of selected area and $C_q$ quantifies its concentration, and systematically compared how these tracers shape our interpretation of galaxy quenching.

Our results demonstrate that different quenching criteria lead to distinct conclusions regarding both quenching modes and their correlations with global galaxy properties:

\begin{itemize}

\item[1.] Quenching mode is more strongly correlated with stellar mass than with halo mass across most diagnostics (Figure~\ref{fig:hmass}). In particular, the fraction of inside-out quenching increases with stellar mass, while outside-in quenching is more frequently observed in lower-mass galaxies, especially satellites. This trend is broadly consistent with theoretical and simulation-based studies, which show that internal processes—such as AGN feedback and morphological quenching—become increasingly important in massive systems, whereas environmental mechanisms preferentially affect lower-mass galaxies.

\item[2.] The dependence of quenching mode on halo mass is weaker and more sensitive to the choice of diagnostic (Figure~\ref{fig:hmass}). While the sSFR-based classification exhibits a clearer halo-mass trend, the D$_n$4000-based results show a weaker or absent dependence. This difference arises from the distinct physical sensitivities of the tracers, including their characteristic timescales (Appendix~\ref{app:sfh_timescales}) and susceptibility to emission-line contamination, rather than reflecting a fundamental inconsistency in the underlying quenching processes.

\item[3.] Different quenching diagnostics probe different stages of galaxy evolution. The sSFR and D$_n$4000 criteria select larger and more diverse galaxy populations spanning a wide range of star formation activity, while the PSB and LI(N)ER criteria identify galaxies in more specific and temporally localized phases associated with recent or late-time quenching (Figure \ref{fig:fullsample}). Because PSB identifies transitional regions, PSB-based quenching patterns should be interpreted as tracing ongoing quenching rather than fully quenched structures.

\item[4.]
We emphasize that even diagnostics with a high degree of overlap can lead to systematically different quenching classifications. In particular, the sSFR and D$_n$4000 criteria select broadly similar quenched regions at both the spaxel and galaxy levels (Figure~\ref{fig:upsetplot}), yet they yield markedly different quenching mode distributions (Figure~\ref{fig:barchart_full}). The sSFR criterion identifies a larger fraction of outside-in systems, whereas the D$_n$4000 criterion more readily classifies galaxies as fully quenched. This discrepancy arises from the distinct limitations and physical sensitivities of the two diagnostics. The sSFR tracer, based on H$\alpha$ emission, is sensitive to ongoing star formation on short timescales, but its reliability can be reduced in regions with weak emission or complex ionization conditions. In contrast, D$_n$4000 traces the cumulative aging of stellar populations and does not rely on emission-line measurements, making it more effective at identifying fully quenched systems within the F$_q$–C$_q$ framework. These results highlight that high overlap between tracers does not guarantee consistent quenching classifications, and careful consideration of tracer-specific limitations is essential when interpreting quenching patterns.

\end{itemize}

In summary, the interpretation of spatially resolved quenching patterns depends strongly on the diagnostic used. While different tracers probe distinct stages of galaxy evolution, they reveal broadly consistent trends—such as the connection between quenching mode and global galaxy properties, particularly stellar mass. At the same time, systematic differences between diagnostics reflect their varying physical sensitivities, timescales, and susceptibility to contamination. A multi-tracer approach is therefore essential not only for identifying robust quenching signatures, but also for understanding the origin of discrepancies and the limitations inherent to each diagnostic.

\color{black}
\acknowledgments
We would like to thank the anonymous referee, who has provided many useful comments and suggestions.
This work is supported by the National Science and Technology Council of Taiwan under the grants NSTC 113-2112-M- 001-006- and NSTC 114-2112-M-001-041-MY3. We thank A. Cooper for providing helpful suggestions to this manuscript.

\appendix

\renewcommand{\thefigure}{A\arabic{figure}}
\renewcommand{\thetable}{A\arabic{table}}
\setcounter{figure}{0}
\setcounter{table}{0}

\section{Illustrative SFH models and temporal sensitivity of quenching criteria}
\begin{figure}[h!]
    \centering
    \includegraphics[width=1\linewidth,height=1\textheight,keepaspectratio]{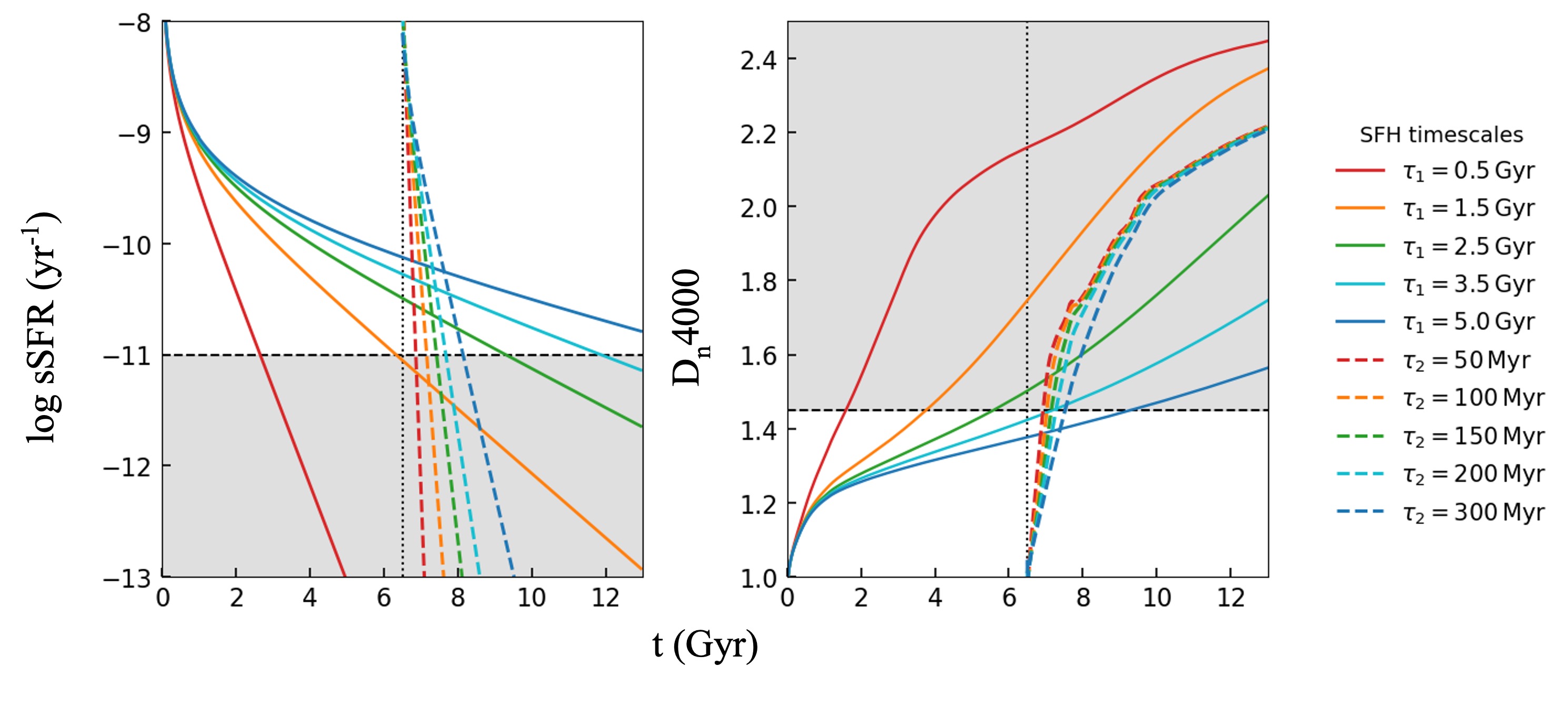}
    \caption{Evolution of sSFR and D$_n$4000 as a function of time for a set of parametric SFHs generated using the Bruzual \& Charlot stellar population synthesis models. The SFHs consist of an initial exponentially declining phase with e-folding times $\tau_1$ = 0.5–5.0 Gyr (solid lines), followed by a rapid quenching phase with e-folding times $\tau_2$ = 50–300 Myr (dashed lines). The left panel shows the evolutionary tracks in the sSFR–time plane, while the right panel presents the corresponding evolution of D$_n$4000. The horizontal dashed line indicates the sSFR threshold adopted in this work ($\log\,\mathrm{sSFR} = -11$), and the horizontal dashed line in the right panel marks the D$_n$4000 threshold of 1.45. The vertical dotted line denotes the onset of the second phase in the adopted double-$\tau$ star formation histories.}
    \label{fig:ssfr_d4000_sfh}
\end{figure}

\begin{figure}[h!]
    \centering
    \includegraphics[width=0.5\linewidth,height=.7\textheight,keepaspectratio]{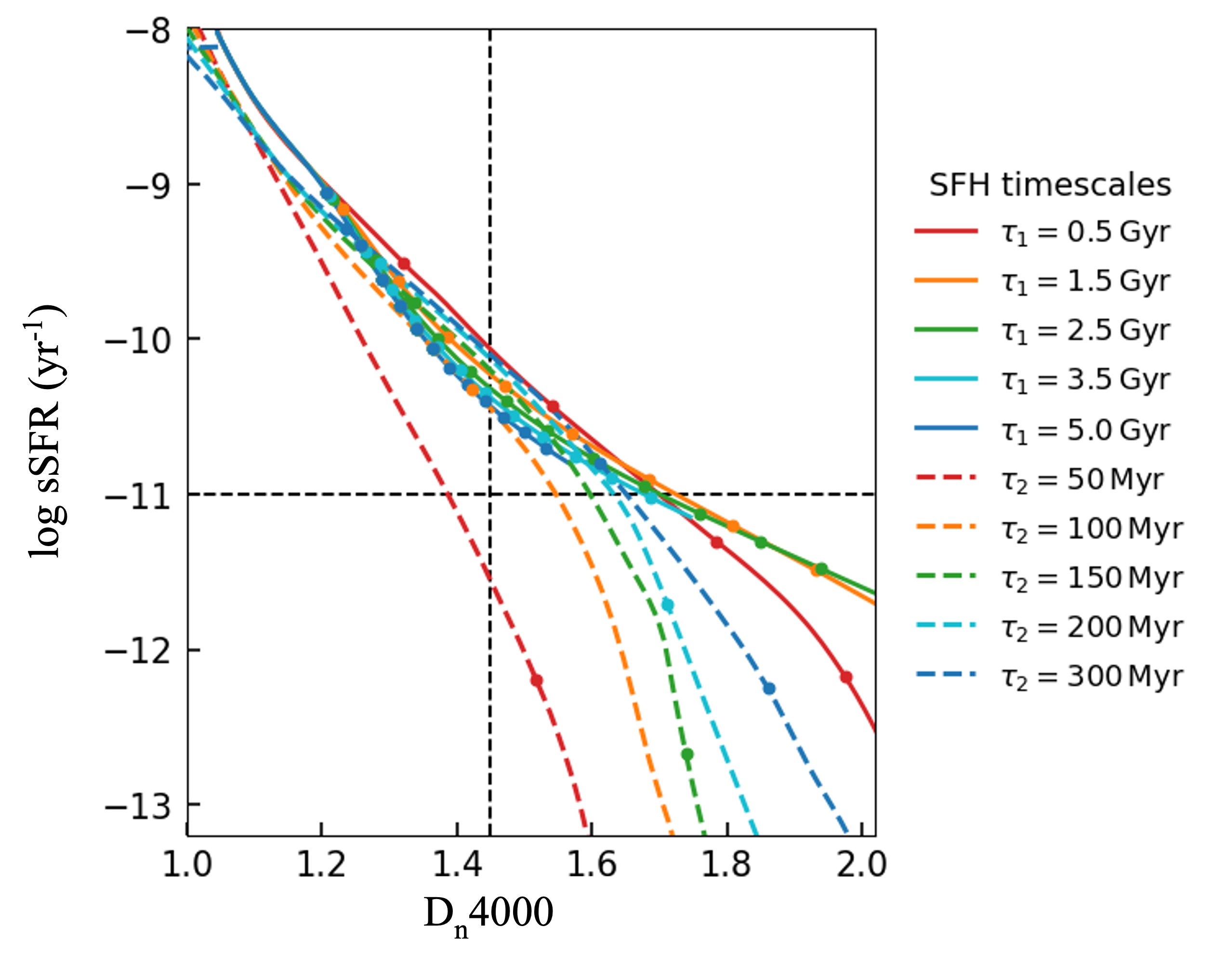}
    \caption{Evolutionary tracks in the sSFR–D$_n$4000 plane for the same set of star formation histories shown in Figure~\ref{fig:ssfr_d4000_sfh}. The colored solid lines represent the slow, exponentially declining phase with e-folding times $\tau_1$ = 0.5–5.0 Gyr, while the dashed lines indicate the subsequent rapid quenching phase with e-folding times $\tau_2$ = 50–300 Myr. The horizontal dashed line marks the adopted sSFR threshold ($\log\,\mathrm{sSFR} = -11$), and the vertical dashed line marks the D$_n$4000 threshold of 1.45.} 
    \label{fig:ssfr_d4000_track}
\end{figure}

\begin{figure}[h!]
    \centering
    \includegraphics[width=0.5\linewidth,height=.7\textheight,keepaspectratio]{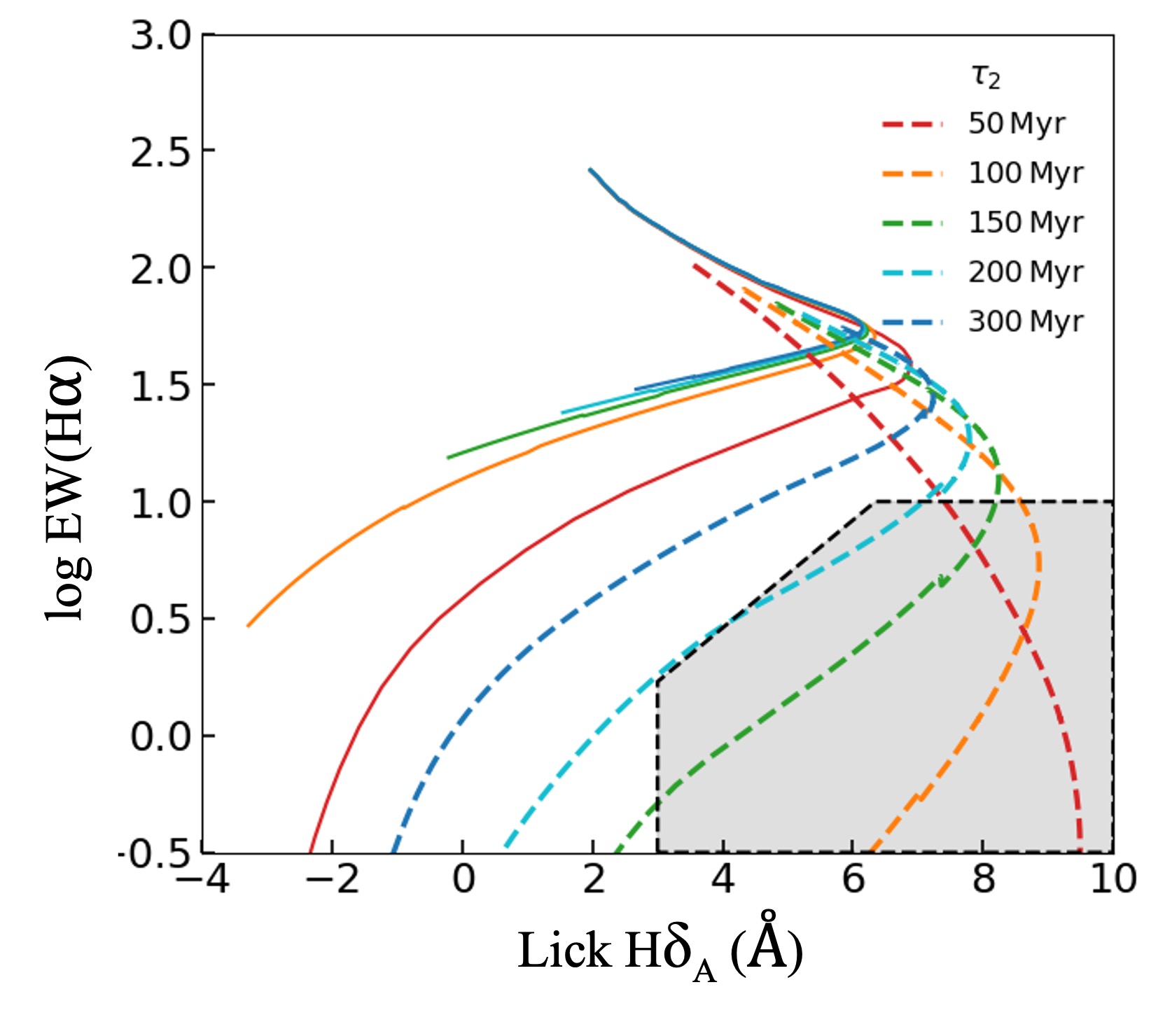}
    \caption{Evolutionary tracks in the EW(H$\alpha$)–H$\delta_\mathrm{A}$ plane for the same set of star formation histories shown in Figures~\ref{fig:ssfr_d4000_sfh} and \ref{fig:ssfr_d4000_track}. The shaded region marks the PSB selection window adopted in this work, defined by strong Balmer absorption and weak nebular emission. Only SFHs undergoing rapid quenching cross the PSB selection region, and they remain within this region for a relatively short time interval.}
    \label{fig:psb_track}
\end{figure}

\label{app:sfh_timescales}

This appendix provides an illustrative interpretation of the characteristic timescales associated with the quenching criteria adopted in this work. Our motivation is to clarify (1) when different criteria are expected to be satisfied along a galaxy evolutionary track, (2) how long a galaxy (or region) can remain within each selection window, and (3) how the diagnostics differ between the timing of threshold crossing and their sensitivity to changes in the quenching timescale. These examples are intended as toy models that qualitatively demonstrate the relative ordering and temporal sensitivity of different diagnostics, rather than to reproduce the full diversity of realistic galaxy star formation histories. These examples are intended as toy models that qualitatively demonstrate the relative ordering and temporal sensitivity of different diagnostics, rather than to reproduce the full diversity of realistic galaxy star formation histories.

\subsection{Toy SFH set following \citet{chen19}}

We construct ten illustrative SFHs. Five are single-exponential models of the form
\[
{\rm SFR}(t)=A\,e^{-t/\tau},
\]
with a normalization $A=1$ (in arbitrary units) and exponential timescales
\[
\tau = 0.5,\ 1.5,\ 2.5,\ 3.5,\ {\rm and}\ 5.0~{\rm Gyr}.
\]

The other five are double-phase models designed to mimic a rapid quenching episode after an earlier declining phase. For $t<t_{\rm burst}$, the SFH follows
\[
{\rm SFR}(t)=A\,e^{-t/\tau_1},
\]
with $\tau_1=5.0~{\rm Gyr}$. At the onset of the second phase, fixed at $t_{\rm burst}=6.5~{\rm Gyr}$, the SFR is reset to an enhanced value and then declines exponentially with a shorter timescale $\tau_2$:
\[
{\rm SFR}(t)= {\rm SFR}_{\rm burst,instant}
\exp\!\left[-(t-t_{\rm burst})/\tau_2\right], \qquad t\ge t_{\rm burst}.
\]
We adopt $\tau_2 = 50,\ 100,\ 150,\ 200,\ {\rm and}\ 300~{\rm Myr}$. The normalization of the second phase is controlled by a burst mass fraction parameter $f_{\rm burst}=0.7$. In the model, the stellar mass formed in the post-burst component is defined as
\[
M_{\rm burst}=\frac{f_{\rm burst}}{1-f_{\rm burst}}\,M_{\rm pre},
\]
where $M_{\rm pre}$ is the stellar mass formed before $t_{\rm burst}$. This choice produces a strong second-phase contribution and is intended only as an illustrative toy model rather than a realistic fit to galaxy SFHs.

The resulting time evolution of sSFR and D$_n$4000 is shown in Figure~\ref{fig:ssfr_d4000_sfh}, where the solid lines represent the single-$\tau$ models and the dashed lines represent the double-phase models. The corresponding evolutionary tracks in the sSFR--D$_n$4000 plane are presented in Figure~\ref{fig:ssfr_d4000_track}, and the evolution of the same models in the EW(H$\alpha$)--H$\delta_{\rm A}$ plane, used to illustrate when the PSB criterion is satisfied, is shown in Figure~\ref{fig:psb_track}.

\subsection{Stellar population synthesis assumptions}

For each SFH, we compute the time evolution of diagnostic quantities using the Bruzual \& Charlot stellar population synthesis models \citep{bc03}, assuming a Chabrier initial mass function \citep{cha03}. 
From the resulting model spectra, we track the evolution of sSFR, the $D_n$4000 index, and the spectral features used in the PSB selection (EW(H$\alpha$) and H$\delta_\mathrm{A}$), measured consistently with the definitions adopted in the main text.

\subsection{Illustrative ordering and time windows of different criteria}

\begin{table*}[ht]
\centering
\caption{Crossing times of different quenching diagnostics for both single-$\tau$ and double-$\tau$ SFH models. Columns $t_{\rm enter}^{\rm sSFR}$ and $t_{\rm enter}^{D_n4000}$ denote the times when each model first satisfies the adopted sSFR and $D_n$4000 thresholds, respectively. For the PSB criterion, $t_{\rm enter}^{\rm PSB}$ and $t_{\rm exit}^{\rm PSB}$ indicate the entry and exit times of the PSB selection window, and $\Delta t_{\rm PSB}$ is the duration spent within this phase. Models that do not enter the PSB region are indicated by dashes.}
\label{tab:sfh_times_all}
\begin{tabular}{lcccccc}
\hline
Model & $t_{\rm enter}^{\rm sSFR}$ & $t_{\rm enter}^{D_n4000}$ & $t_{\rm enter}^{\rm PSB}$ & $t_{\rm exit}^{\rm PSB}$ & $\Delta t_{\rm PSB}$ \\
 & (Gyr) & (Gyr) & (Gyr) & (Gyr) & (Gyr) \\
\hline
single $\tau=0.5$ Gyr & 2.70 & 1.60 & --- & --- & --- \\
single $\tau=1.5$ Gyr & 6.32 & 3.80 & --- & --- & --- \\
single $\tau=2.5$ Gyr & 9.30 & 5.60 & --- & --- & --- \\
single $\tau=3.5$ Gyr & 11.90 & 7.20 & --- & --- & --- \\
single $\tau=5.0$ Gyr & --- & 9.35 & --- & --- & --- \\
\hline
double $\tau_2=50$ Myr  & 6.86 & 6.93 & 6.72 & 6.93 & 0.20 \\
double $\tau_2=100$ Myr & 7.16 & 7.03 & 6.92 & 7.35 & 0.43 \\
double $\tau_2=150$ Myr & 7.45 & 7.15 & 7.12 & 7.70 & 0.58 \\
double $\tau_2=200$ Myr & 7.70 & 7.27 & 7.40 & 7.65 & 0.25 \\
double $\tau_2=300$ Myr & 8.15 & 7.55 & --- & --- & --- \\
\hline
\end{tabular}
\end{table*}

Figures~\ref{fig:ssfr_d4000_sfh} and \ref{fig:ssfr_d4000_track} illustrate the evolution of sSFR and $D_n$4000 for the ten SFHs. We focus on the evolution following the onset of quenching (vertical dashed line), and track how each diagnostic responds as a function of time since quenching.

We find that the relative ordering of the threshold crossings depends on the adopted diagnostic definitions. In most of the models considered here, the $D_n$4000 threshold ($D_n4000 \gtrsim 1.45$) is reached earlier than the sSFR threshold, as quantified in Table~\ref{tab:sfh_times_all}. This indicates that, under our adopted criteria, $D_n$4000 can identify quenched regions at earlier stages following the suppression of star formation, while the sSFR threshold is typically reached at later times.

At the same time, the two diagnostics differ in their sensitivity to the quenching timescale. The sSFR threshold depends strongly on the rate of star formation decline: more rapid quenching (shorter $\tau$ or $\tau_2$) leads to substantially earlier crossings, while more gradual quenching delays the transition. In contrast, the evolution of $D_n$4000 depends more weakly on the quenching timescale and is not strictly monotonic, since it reflects the cumulative aging of the stellar population rather than the instantaneous suppression of star formation.

This difference explains why, in the single-$\tau$ models, larger $\tau$ values can still correspond to earlier crossings of the $D_n$4000 threshold, despite a slower decline in star formation. Overall, sSFR is more sensitive to variations in the quenching timescale, whereas $D_n$4000 traces the integrated stellar population history and evolves more gradually.

In contrast, the PSB selection is only satisfied in models with rapid quenching and over a limited time interval. The PSB phase is typically entered shortly after the onset of rapid quenching, but not necessarily earlier than the sSFR threshold, and persists for only $\sim$0.2--0.6~Gyr. None of the single-$\tau$ models enter the PSB selection window, further emphasizing that the PSB criterion selectively traces short-lived phases associated with rapid quenching rather than gradual declines in star formation.

Figure~\ref{fig:psb_track} shows the evolution of the same SFHs in the EW(H$\alpha$)--H$\delta_\mathrm{A}$ plane. Only rapidly quenched SFHs cross the PSB selection region, and they remain within this region for a relatively short duration, consistent with the transient nature of the PSB phase.

Taken together, these toy models demonstrate that the sSFR, $D_n$4000, and PSB criteria probe distinct temporal aspects of the quenching process. In particular, the time at which a diagnostic first reaches its adopted threshold is not the same as its sensitivity to changes in the quenching timescale. The sSFR is highly sensitive to the rate of star formation suppression, whereas $D_n$4000 reflects the cumulative stellar population age and therefore evolves more gradually, even though under the adopted thresholds it can reach the quenched selection boundary at relatively early stages. The PSB criterion, by contrast, traces a short-lived phase associated with rapid quenching.

\color{black}

\end{document}